# Novel approaches to urban science problems: human mobility description by physical analogy of electric circuit network based on GPS data

**Authors:** Zhihua Zhong[1], Hideki Tayakasu[1,2], Misako Takayasu[1]

**Institutions and Affiliations:**
[1] *School of Computing, Tokyo Institute of Technology, Tokyo, Japan.*
[2] *Sony Computer Science Laboratories, Tokyo, Japan.*

**Corresponding Author:**
Misako Takayasu
School of Computing, Tokyo Institute of Technology, 2-12-1 Ookayama, Meguro-ku, Tokyo, Japan, 152-8550
Email: takayasu.m.aa@m.titech.ac.jp

**Abstract**

Human mobility in an urban area is complicated; the origins, destinations, and transport methods of each person differ. The quantitative description of urban human mobility has recently attracted the attention of researchers, and it highly related to urban science problems. Herein, combined with physics inspiration, we introduce a revised electric circuit model (RECM) in which moving people are regarded as charged particles and analogical concepts of electromagnetism such as human conductivity and human potential enable us to capture the characteristics of urban human mobility. We introduce the unit system, ensure the uniqueness of the calculation result, and reduce the computation cost of the algorithm to 1/10000 compared with the original ECM, making the model more universal and easier to use. We compared features including human conductivity and potential between different cities in Japan to show our improvement of the universality and the application range of the model. Furthermore, based on inspiration of physics, we propose a route generation model (RGM) to simulate a human flow pattern that automatically determines suitable routes between a given origin and destination as a source and sink, respectively. These discoveries are expected to lead to new approaches to the solution of urban science problems.

**1. Introduction**

Human mobility affects basic aspects of human daily life in an urban city: economic development, energy use, and evolutions of the shape and structure of cities[1,2,3,4,5,6,7,8,9,10,11,12,13,14]. It is essential to examine traces of human activities in urban areas to control the spread of infectious diseases such as COVID-19[15,16].

Previous studies on human mobility can be roughly categorized into two types: macroscopic and microscopic. Macroscopic research mostly focuses on non-trivial macro relationships of human mobility between cities or countries through models, such as the gravity model and the radius model[17,18,19,20,21,22,23]. In these models, the target of the study is the flow of people moving between two separate cities A and B. The intensity of this flow is approximated using a function, which is the product of the populations in these cities divided by the power law of the distance between those cities. This function, is based on a generalized form of Newton's law of gravitation for the attractive force between two planets. This simple model can quantitatively describe the human flow between cities; however, it neglects most of the information on individual mobility within the cities. Regarding microscopic research, it has applied machine-learning algorithms to big GPS data with the primary aim of examining the detailed moving trajectories of individuals to predict their individualized movements within a city[24,25,26,27,28,29,30,31].

Between macroscopic and microscopic perspective, there are studies on the mesoscopic properties of human mobility, which ignore individual movements and instead focus on the flow of people within small areas in big cities[32,33,34]. Recently, Shida et al.[35] proposed a novel model of human mobility, called the electric circuit model (ECM), which is inspired by the physical structure of an electric circuit, to parse time-dependent human flow patterns in urban areas. The basic idea of the ECM is to treat each moving person as a positively charged particle and describe massive human flow as electric currents on an imaginary square lattice of an electric circuit network defined over an entire city. The values of conductivity for resistors in the circuit area, which are defined from given GPS data, can clearly describe the type of infrastructure in a city; for example, conductivities along major railways are high to support huge transport system. During rush hours in the morning and evening, strong human electric potential of opposite signs appear in the city center, whereas at noon, the currents more closely resemble random thermal noises. Although the ECM was expected to be widely applicable, it has two major shortcomings: the definition of the electric circuit is not universal and depends on the given data; the computational cost associated with resistors, which depend on the GPS data, are large and not scalable for large GPS data; and the calculation result cannot be compared with different cities' or countries' due to that it changes with the hyperparameter of optimization algorithms.

Therefore, we revised the ECM, introducing universal definitions, such as human ampere, human ohm, and human voltage, which can be applied to any type of GPS data in a city. And we proposed a more universal definition for human resistance to eliminate limitations of ECM. The revised electric circuit model (RECM) drastically reduces the computational costs to 1/10000, secures uniqueness in the calculated results, enabling comparisons on the GPS data of different cities even different countries when data are available, and significantly extends the application range of the existing ECM. We compared the human conductivity and potential, reflecting infrastructure level and human movement tendency respectively, between different cities to show our model improvement. Additionally, based on RECM we proposed a new model called route generation model (RGM), which generates potential routes when provided with the origin and destination in situations with an accident or disaster area and no available transport.

Our discoveries provide new approaches to capture the characteristics of human mobility pattern in urban cities. In the future, further research related to route recommendation, natural disaster simulation, and urban design, such as research on human mobility simulation for constructing new public transport systems, can be conducted.

## 2. Results

This section describes the performance of our models, using the greater Tokyo area, which is one of the largest metropolitan cities in the world and contains one-fourth of the Japanese population (Data description shown in Method 4.1). Section 2.1 introduced the inspiration of human current from physics. Section 2.2 presents the results of the improvements accompanying the RECM. Section 2.3 compares the human conductivity and potential between other different cities in Japan. Finally, section 2.4 describes the performance of the RGM and a proposed application to simulate human flow patterns in a situation where traffic is blocked in a part of the city owing to an accident or disaster.

### 2.1 Observation of human current

In electromagnetics, a current $i$ through a cross-sectional area $s$ of a metal conductor is given by $i = -espv$, where $p$ is the number density of the moving electrons with charge $-e$, and $v$ is the mean velocity component perpendicular to the cross-section. Therefore, our basic idea is to consider that moving people, carrying positive unit charges, forms human current. We defined the human current through a cross-sectional line of length $S$ as follows (see Fig. 1 (a)):

$$I = SPV/c \qquad (1)$$

where $P$ is the population density of moving people observed in the GPS data; $V$ is the mean velocity component perpendicular to the line averaged over the moving people; and $c$ is the coverage of the GPS data that was defined as the number of users in the database $N_{data} = 2,957,390$ divided by the total population of Japan $N_{total} = 124,490,000$ in 2022. Equation (1) represents the estimated total number of people, including non-GPS users, moving through the cross-sectional line in unit time estimated from the observable GPS users. We measured the length in terms of [km] and the speed in terms of [km/h]; so the unit of the number density is [1/km$^2$] and that of the human current is [1/h], which we referred to as a human ampere [HA].

For the subsequent data analysis, the city map was divided into square meshes with length $S = 0.5$ [km], as shown in Fig. 1 (b). Human current was defined on the links connecting the centers of neighboring meshes (Fig. 1 (c)), and correspondingly, we introduced an electrical circuit network covering the whole city, as shown in Fig. 1 (d-e). Detailed calculation processes are shown in section Method; notation used in this study will be introduced in section 4.2; and main results will be shown in the following section.

**2.2 Improvement in the RECM**

By calculating the population time series over one week at typical locations with high (blue), medium (orange), and low (green) populations, Fig. 1 (f) shows clear circadian rhythms representative of people in morning and evening rushes, except on weekends. The mean velocity time series of the people appeared in these three locations is shown in Fig. 1 (g). In the case of the high population mesh, which is located in the city center, the mean velocities were small because the directions of movements were scattered; in the case of the medium population mesh, which is located in a suburban residential area, the velocity profile showed a clear circadian rhythm, owing to people moving toward the city center in the morning and in the reverse direction in the evening. As defined in Method 4.3 equation (2), the human currents were calculated as the average number of products of population and mean velocity in two adjacent meshes, as shown in Fig. 1 (h). Strong currents and periodicity were confirmed in the medium population mesh, especially on the mornings of weekdays.

Fig. 1 (i) shows the distribution of human current for 4 different places every 0.5 h from 5 a.m. to 24 p.m. for weekdays in 2022. The positive and negative values were both plotted separately and also together for i-1, i-2, and i-3 in the semi-log scale. In each case, the plot was approximated using an exponential distribution. Fig. 1 (i-4) shows the case where the distribution of currents for the north-south and east-west directions followed exponential distributions with different mean values owing to the direction of the railway.

As mentioned in introduction, to solve the limitations of computational cost and not unique calculation result of resistance and potential, we proposed a new definition for human resistance, as defined in Method 4.3 equation (4), to solve these two drawbacks simultaneously (details improvement about time complexity discussed in section Method 4.5; since it was not treated as optimization problem anymore that results are sensitive to initial value and hyperparameter, uniqueness of calculation was achieved, whose details will be discussed in Supplementary Material 2). Furthermore, quantities with unclear practical meaning calculated by the black-box optimization algorithms in the previous research can be explained clearly in our study after we parsed its mathematical formula and built up its mathematical foundation.

Human conductance (equation (5)), $\rho_{(d,t,L,\sigma)}$, which is the inverse value of human resistance, is the slope of dashed lines in Fig. 1 (i). In Fig. 2 (a-b), the spatial distribution of human conductance is high near the railway and high-speed road, while it is low near the mountainous and rural areas. Human resistance and human conductivity reflect the ease of human flow and indicate the infrastructure level of a location, as discussed in Method 4.3.

As defined in equation (6), human voltage characterizes the amount of human current normalized by the average of absolute current at the same location. As known from the distributions in Fig. 1 (i), human voltage typically takes a value around 0, and the range of value is practically between [-10, 10]. As shown in Fig. 2 (c), human voltage time series for a week showed that the maximum voltage had been approximately 4 in the mesh including Nagatsuda Station (living area) on Monday located in a suburb, implying that the current value had been approximately four times the average absolute current at that station during rush hour.

Moreover, as defined in equation (7), human potential showed that human behaviour resembles that of charged particles, which tend to move from a higher potential area to a lower potential area. Fig. 2 (e-g) shows the spatial pattern of human potential in the morning rush hour (7:30–8:00), noon (13:30–14:00), and evening (18:30–19:00) respectively, which also meant that we had also successfully reproduced the human potential spatial distribution after changing the definition of human resistance (Detailed discussion in Supplementary Material 2).

We can learn the tendency of how people move and distinguish where are sink and source in an urban city by observing human potential in different time period. In Fig. 2 (h), the human potential time series within a day at the Tokyo (blue), Nagatsuda (orange), and Kazo (green) stations showed that Tokyo Station, which has many business and commercial areas nearby, was a sink in the morning and a source in the evening. In contrast, Nagatsuda Station, which has many nearby residential areas, was a source in the morning and a sink in the evening. Since the road near Kazo Station had normal traffic flows and no residents or business areas nearby, people tended to pass through it, resulting in the human potential being close to zero all day.

In Fig. 2 (d), electrical energy dissipation, as defined in equation (8), shows that in the morning rush hour, high energy dissipation meshes shown in red are not in the city center but located around the suburban areas indicating that large accumulation of people occurs in these areas during morning rush hour.

Overall, human behavior in an urban city can be characterized by analogized concepts of electromagnetics, such as human current [HA], resistance [HΩ], and potential [HV]. And based on these quantities, auto classification of areas in the urban city by using unsupervised algorithm such as k-means clustering can be conducted in the future research, since time series pattern of them are quite different between resident, business, commercial, city boundary, and rural areas. Furthermore, since we made the application range of ECM more universal by securing the uniqueness of the calculation result, comparison of these quantities between different countries will also be interesting as long as other researchers have the corresponding GPS data in other countries.

**2.3 Comparison between different cities**

The calculation result of the former ECM depends on the fine-tuning of the parameter of ADAM, so the result of human resistance varies and cannot be directly compared between different cities. Since the calculation of human voltage, charge, and potential is based on the resistance value, the human potential value was also parameter-dependent, and not unique. In other words, the former ECM could only be used once in one city, and the application range was limited. Thanks to the improvement of our model by changing the calculation method of the resistance value, we ensured the uniqueness of the calculation result, which enabled us to compare the characteristics in RECM between different cities in Japan.

As shown in Fig. 3 (a), the bar plot of population and maximum human conductivity between 5 cities, including Tokyo, Osaka, Kyoto, Sendai, and Kumamoto, shows that the infrastructure level of a city grows with the population. When more workers enter a city, the population density will increase, and better infrastructure will need to be built to support the human movement of many people. Fig. 3 (b) shows the linear relation between population and maximum conductivity (best infrastructure level) of a city; linear regression was used to determine the slope, $k = 0.0007$, which means that per 1400 people, on average, 1 unit of maximum conductivity increases.

Fig. 3 (c) shows the cumulative distribution function of human resistance for the 6 cities in log-log scale, and we found that these distributions are approximated by a power law with an exponent of about -0.7. Fig.3 (d) shows the distribution of absolute values of human potential for all times and places in each city in the morning rush hour from 7:30 am to 8:00 am. Tokyo, as Japan's largest metropolitan area, has the greatest potential value; followed by Sapporo and Osaka, and then, Kumamoto, Sendai and Kyoto.

Fig. 3 (e) compares the maximum and minimum values of the potential in each city observed in 3 different time zones, morning, noon and evening. For all cities, the difference between the maximum and minimum potential values in the morning rush hour is about twice as large as that in the evening rush hour, because the time to return home in the evening has a wider range than the time to arrive at work in the morning, and major workplaces are concentrated in the city center, but residences are dispersed throughout the area, so the concentration of people flows is very different in the morning and in the evening. On the other hand, there is no clear directional flow at noon, the human flow pattern is close to random noise, and the maximum and minimum potential values are close to zero in all cities.

We now define the center of human potential for each city as the place with the minimum potential value in the morning rush hour, and then define the effective radius of the city as the average of the minimum and maximum lengths from the potential center to the boundary with zero potential. It turns out that the radius is approximately proportional to the range of the potential difference in the morning rush hour, namely, the highest voltage minus the lowest, as shown in Fig.3 (f). This result is consistent with the intuition that the range of the potential difference is given by the path integral from the lowest point to the highest point, and if the slope of the local potential is roughly constant, then the effective radius of the city determines the maximum potential difference.

**2.4 Route generation model**

In the RECM, temporal human potential can describe the tendency of the people in an urban city to exhibit macro movements. However, it cannot be used to generate a microscopic movement route for each individual. Herein, based on RECM, we propose a new model, RGM, that can generate possible routes for people when given specific origins and destinations. Based on the physical architecture of the electric circuit, the flow pattern of human current from the origin to the destination can be obtained by solving Kirchhoff's law[36] under the condition that electrodes are placed at the origin and destination points.

As shown in Fig. 4 (a), Ayase Station and Shinanomachi Station in Tokyo were set as the origin and destination, respectively. A unit voltage was applied between these points, which were approximately 25 km apart, in a straight line. Then, the human current pattern was calculated. From the result, the most popular route (as indicated by the red line), was constructed by connecting the largest current flow direction at each mesh from the origin based on the greedy algorithm. Fig. 4 (b) shows the recommended route from Ayase Station and Shinanomachi Station provided by Google Maps. Not only the largest flux route, but also the second and third largest flux routes as green, yellow and yellow-green lines could be found; interestingly, the routes generated by RGM were consistent with the real-world paths suggested by Google Maps. Unlike traditional shortest-path algorithms, such as A* and Dijkstra[37], the algorithms used in Google Maps that can only generate a single path from the origin to the destination to find a local solution, our model uses the global information of the system to generate all the possible paths in an electric circuit network. Therefore, ideas such as route recommendation through the development of software or smartphone applications that merge our model with traditional algorithms used in Google Maps can also be considered worthy of further research (other examples of route generation are shown in Supplementary Material 3).

Another application of our model is to predict how people will move in the event of a disaster. We assumed a situation where the trains around Tokyo Station were not operating due to a disaster such as a power outage or an earthquake. For this simulation, the conductivity of the link in the disaster area was set to 0 (black rectangle), as shown in Fig. 4 (c), and the model successfully provided solutions for any given origin or destination. In the case where the origin and destination were the same as those in Fig. 4 (a), the most popular route was indicated by the red line, which coincided with the third most popular route to avoide the disaster area. In this way, the simulation of human flow patterns before the occurrence of natural disasters or accidents for better preparation is also an interesting and important future application for the government.

**3. Discussion and conclusion**

In this study, we showed that human behavior resembled electrically charged particle's behavior and that the RECM was a powerful tool for research on human mobility and urban science. We introduced universal definitions for the existing ECM and extended its application range, as discussed in section 2.1 and 2.2; compared human conductivity and potential between different cities in section 2.3; and proposed a new model RGM to realize the generation of possible detailed routes between a given origin and destination, as detailed in Section 2.4.

Our model has some limitations. First, the RGM does not consider time, and the best routes to travel to a particular destination may differ over time; therefore, the accuracy of the model may differ when time is considered. Second, the model does not take into account competitive relationships such as the following. If everyone uses their mobile phones to check the best route and follows the same route, traffic congestion will occur, making the recommended route inappropriate. So, competitive relationships need to be considered.

Regarding future research, the applications of our proposed models related to urban design can be very wide. For example, when building a new railway or station, changes in human mobility can be assessed by reducing the resistance value of links where new railway is planned and solving RGM to simulate changes in human mobility to quantitatively evaluate the effect of adding or removing some infrastructure at certain locations.

As shown in Fig. 4 (c), when a disaster occurs near Tokyo Station and people are temporarily unable to pass through, we learned that people would use another transportation option, i.e., the Yamanote Line (Route No. 3), to bypass the area affected by the disaster. In this way, further research on the natural disaster simulator would be both interesting and valuable[38,39,40,41,42]. Importantly, if the government is well informed about how people move during natural disasters, it can make more effective preparations, such as building appropriate infrastructure in specific areas prone to such disasters.

Overall, combined with the inspiration of physics, we built up new models, promising to evolve to the means to solve urban science problems. We hope that our contribution can be used by other researches, such as comparing different human flow patterns between different counties by using our model.

## 4. Methods

In this section, we summarize the information that is necessary to implement our methods for those researchers who have access to individual GPS data. Section 4.1 introduces the data we used in this study, and any researcher who has GPS data fulfils the format of the GPS data example shown in Supplementary Material 1 can implement our models to conduct further research immediately. Section 4.2 includes the notation used in this study. Section 4.3 describes the theoretical definition for quantities, analogical concept from electromagnetics, used in this study. Section 4.4 illustrates the detailed procedure for practical computation, including data preprocessing (**Step 1**), calculations for the RECM (**Step 2**), and calculations for the RGM (**Step 3**). Finally, section 4.5 simply analyses the improvement of the time complexity of the algorithms, and shows that in terms of numerical calculation, it is feasible to deploy to the practical problem.

### 4.1 Data source.

A Japanese company, Agoop Corp., provided the database used in this study (an example of the data can be found in Supplementary 1). It contains detailed GPS data (including parameters such as user ID, time, longitude, latitude, speed, and angle) collected from over one million smartphone users in Japan from 2022/3 to 2022/11. We focused on the weekday data because the human mobility patterns differed on weekends, as previously discussed in result section.

For privacy protection reasons, the GPS data used in this research were obtained with the consent of smartphone application users. These data were anonymized in advance in accordance with the guidelines set by the general incorporated association Location Based Marketing Association (LBMA) regarding the use of location information data in Japan. Since information such as the name, address, gender, and age of a user was not included, and each user ID renewed at midnight, it was infeasible to identify individuals from just their user attributes or track their cyclical movements across days.

Overall, no privacy issues were faced when conducting this study. Approximately one billion pieces of data were available daily, with the database totalling a size of 22 TB for 2022. Regarding the computational environment, the CPU was 2 × Intel Xeon Gold 6248R (24 core); the memory was 500 GB; and the OS was Linux Ubuntu.

## 4.2 Notation

As shown in Fig. 1 (e), each node of the electric circuit network has four links pointing in the right (east), up (north), left (west), and down (south) directions. In the system, the location of each node is represented by a set of integers $(x, y)$, $x, y \in N$. To express directions, we defined a set of linear maps as $\sigma \in \{\sigma_{+x}, \sigma_{+y}, \sigma_{-x}, \sigma_{-y}\}$, where $\sigma_{+x}: (x, y) \to (x + 1, y)$, $\sigma_{+y}: (x, y) \to (x, y + 1)$, $\sigma_{-x}: (x, y) \to (x - 1, y)$, and $\sigma_{-y}: (x, y) \to (x, y - 1)$. Therefore, $\sigma_{+x}\sigma_{-x} = 1$, $\sigma_{+y}\sigma_{-y} = 1$, and $-\sigma := \sigma^{-1} \in \{\sigma_{-x}, \sigma_{-y}, \sigma_{+x}, \sigma_{+y}\}$ (inverse mapping). The terms $\sigma_{+x}, \sigma_{+y}, \sigma_{-x}$, and $\sigma_{-y}$ expressed the directions right, up, left, and down, respectively. Furthermore, $\sigma_0 = 1$, $\sigma_1 = \sigma_{+x}$, $\sigma_2 = \sigma_{+y}$, $\sigma_3 = \sigma_{-x}$, and , $\sigma_4 = \sigma_{-y}$ were used for expressing rotations, which are discussed later. For example, $I_{(d,t,L,\sigma)}$ is the human current value on day d, at time period t, in location $L$ ($L = (x, y)$), and in direction $\sigma$ (note that the current is defined on the link connecting node $L$ and the neighboring node $\sigma(L)$).

## 4.3 Definition of variables
**Human current**

According to the architecture of the lattice electric circuit network shown in Fig. 1 (d), there are four link pointing to four different direction for a node. Human current $I_{(d,t,L,\sigma)}$ of the link pointing in the right direction can be defined by the following equation, which is equivalent to equation (1) (the basic inspiration and idea we mentioned in section 2.1), averaged over two neighboring meshes in the pointing direction:

$$I_{(d,t,L,\sigma_{+x})} = \frac{(v_{(d,t,L,\sigma_{+x})}*p_{(d,t,L)} + v_{(d,t,\sigma_{+x}(L),\sigma_{+x})}*p_{(d,t,\sigma_{+x}(L))})}{2} * \frac{1}{c} \quad [HA] \quad (2)$$

where $p_{(d,t,L)}$ denotes the density of people with non-zero velocities, who appeared in the mesh of location $L$, at day $d$, in time period $[t, t+0.5h]$. From the data, the population density was calculated by multiplying the number of moving people (with non-zero speeds) in each mesh every 30 min with the normalization factor $1/S^2$, where $S = 0.5\ km$, the length of mesh. $v_{(d,t,L,\sigma_{+x})}$ is the mean value of the $\sigma_{+x}$ direction components (pointing right) of the velocities of those people in the same location and time period. Human currents in the up direction can be calculated by replacing $\sigma_{+x}$ to $\sigma_{+y}$; and other directions can be similarly defined. In the study, current was divided by coverage to proportionally quantify the flow of people that could not be directly observed. Following the definition of current, the following relationship holds between neighboring currents in opposite direction:

$$I_{(d,t,L,\sigma)} = -I_{(d,t,\sigma(L),-\sigma)} \quad (3)$$

**Human resistance**

Regarding to human resistance $R_{(L,\sigma)}$ [HΩ], which take time-invariant values reflecting the transportation infrastructure of each place, in the previous study[35], resistances were determined such that the rotation of the static electric field would be close to 0, which is the condition fulfilled by a static electric circuit. However, in our study, the values of the resistances were found to be proportional to the inverse of the maximum absolute value of currents. Since the maximum was not a statistically stable quantity, we checked several candidates, as discussed in Supplementary Material 2, and concluded that the following definition of human resistance would be much easier to calculate from the data and interpret, and that approximately rotation-free electric fields, fluctuation dissipation theorem, and human potential spatial distribution could also be reproduced:

$$R_{(L,\sigma)} = \frac{1}{\operatorname*{mean}_{d,t}\{|I_{(d,t,L,\sigma)}|\}} \quad [H\Omega] \tag{4}$$

**Human conductivity**

Human conductivity is simply defined as the inverse of the resistance by the following formula:

$$\rho_{(L,\sigma)} = \frac{1}{R_{(L,\sigma)}} = \operatorname*{mean}_{d,t}\{|I_{(d,t,L,\sigma)}|\} \tag{5}$$

By combining equation (2) and (5), it is easy to show that when population $p_{(d,t,L)}$ is fixed, $\rho_{(d,t,L,\sigma)}$ will only increase when $v_{(d,t,L,\sigma)}$ increases (faster transportation). On the contrary, when the velocity $v_{(d,t,L,\sigma)}$ is fixed, $\rho_{(d,t,L,\sigma)}$ will only increase when $p_{(d,t,L)}$ increases (the capability to transport people simultaneously at a given moment). Therefore, human conductivity and resistance reflects infrastructure level of a place in a city.

**Human voltage**

According to Ohm's law, human voltage $E_{(d,t,L,\sigma)}$ [HV] on a link is simply defined by the following formula as an analogical concept:

$$E_{(d,t,L,\sigma)} = I_{(d,t,L,\sigma)} * R_{(L,\sigma)} = \frac{I_{(d,t,L,\sigma)}}{\operatorname*{mean}_{d,t}\{|I_{(d,t,L,\sigma)}|\}} \quad [HV] \tag{6}$$

**Temporal human potential**

According to Helmholtz's theorem, given the spatial distribution of human voltages at a given time period, the vector field of the electric field can be decomposed into a scalar potential component without rotation and a vector potential component without divergence. In this research, the temporal human potential $\emptyset_{(d,t,L)}$ [HV] on the node of all meshes was defined by solving the Poisson equation in a discrete case as follows:

$$\sum_j \Delta_{ij} \cdot \emptyset_{(d,t,L=j)} = div(E_{(d,t,L=j)}) \tag{7}$$

where $\Delta_{ij}$ is the Laplacian matrix[43,44,45,46,47] of the electric circuit network, and human electric charge can be calculated by the divergence of voltage $div(E_{(d,t,L=j)})$.

**Electrical energy dissipation**

According to Joule's Law, electrical energy dissipation is defined using the following formula as an analogical concept:

$$W_{(d,t,L,\sigma)} = I_{(d,t,L,\sigma)}^2 * R_{(L,\sigma)} = \frac{I_{(d,t,L,\sigma)}^2}{\operatorname*{mean}_{d,t}\{|I_{(d,t,L,\sigma)}|\}} \tag{8}$$

### 4.4 Detailed calculation procedure

**Step 1 (Data preprocessing part):**

**Step 1-1.** Determine the observation day *d* (neglect weekends), divide a day into 38 time periods *t* per 30 min from 5 a.m. to 24 p.m. (for example, 5:00–5:30, 5:30–6:00, …, 23:30–24:00), and remove GPS data records with a velocity of zero or missing value (in this study, day *d* was from March 2022 to November 2022)

**Step 1-2.** Cut the map into meshes, each with a size of 0.5 [km] * 0.5 [km], and tag the ID for each node (denote node ID as $L$, $L \in \{(0,0),\ (0,1), (1,0), \ldots, (i,j)\}$).

**Step 1-3.** Define $\sigma_{+x}$ as a function to input a node ID and return the ID of the node that is on the right of that node. Similarly, also define $\sigma_{+y}, \sigma_{-x}$ and $\sigma_{-y}$ (denote the direction mapping set as $\sigma$, $\sigma \in \{\sigma_{+x}, \sigma_{+y}, \sigma_{-x}, \sigma_{-y}\}$)

**Step 2 (RECM part):**

**Step 2-1.** Calculate the population (number of people) $p_{(d,t,L)}$ comprising people with non-zero velocities and the instant velocity vector for each person ($i$) $v_{(d,t,L)}^{(i)}$ using the speed and course (angle) variables in the raw GPS data for every node. (If there are no instant speed and course variables in raw data, speed and angle can also be deduced by (user, longitude, latitude, time) log)

**Step 2-2.** Calculate the mean velocity of each person $v_{(d,t,L)}$ for each node, where $v_{(d,t,L)} = \text{mean}_i \{v_{(d,t,L)}^{(i)}\}$. Decompose the vector $v_{(d,t,L)}$ into its +x, -x, +y, and -y components $v_{(d,t,L,\sigma_{+x})}$, $v_{(d,t,L,\sigma_{-x})}$, $v_{(d,t,L,\sigma_{+y})}$ and $v_{(d,t,L,\sigma_{-y})}$ respectively.

**Step 2-3.** Calculation of human current: given a node, there are four directions to adjacent nodes, $I_{(d,t,L,\sigma_{+x})}$, $I_{(d,t,L,\sigma_{+y})}$, $I_{(d,t,L,\sigma_{-x})}$, and $I_{(d,t,L,\sigma_{-y})}$, which represent right, up, left, and down currents, respectively. Traverse all the nodes $L$, days $d$, and time periods $t$, and calculate the current using equation (2) (time complexity: $O(d*t*L)$).

**Step 2-4.** Calculation of human resistance: Traverse all the nodes $L$, days $d$, and time periods $t$ and calculate the resistance using equation (4) (time complexity of the new method: $O(d*t*L)$).

**Step 2-5.** Calculation of human potential: Calculate the charge $Q_{(d,t,L)}$, the divergence of human voltage, at date $d$, time period $t$ and location $L$ using the following equation:

$$Q_{(d,t,L)} = div(I_{d,t,L} * R_L) = \sum_{i=1}^{4} I_{(d,t,L,\sigma_i)} * R_{(L,\sigma)} \qquad (9)$$

The structure of the electrical circuit network on the map is fixed when the observed data are provided. Find the largest connected component of the network using the breadth-first search (BFS) algorithm. The adjacent matrix is defined as follows:

$$A_{i,j} = \begin{cases} 1, & \text{node } i \text{ is adjacent to } j \\ 0, & \text{else} \end{cases} \qquad (10)$$

Before to define diagonal matrix, sea area and land area need to be defined. According to data downloaded from "Ministry of Land, Infrastructure, Transport and Tourism" Japan, the ratio of sea for each mesh can be calculated, and the sea ratio above 90% of a mesh will be defined as sea area, otherwise it will be defined as land area. Sea nodes will be removed before the calculation, land node adjacent to sea node will be called as sea boundary. Then diagonal matrix for the node on the sea boundary can be defined as follows:

$$D_{i,j} = \begin{cases} \sum_j A_{i,j}, & \text{case 1} \\ 4, & \text{case 2} \\ 0, & \text{case 3} \end{cases} \qquad (11)$$

where $\text{case 1}: i = j \text{ and node } i \text{ is adjacent to the sea area}$
$\text{case 2}: i = j \text{ and node } i \text{ is on the other area}$
$\text{case 3}: i \neq j$

Then, solve the human potential $\emptyset_{d,t,l}$ at day $d$, time period $t$, and location $L$ as follows (More details show in Supplementary Material 4):

$$\left(D_{i,j} - A_{i,j}\right) * \begin{pmatrix} \emptyset_{d,t,1} \\ \dots \\ \emptyset_{d,t,L} \end{pmatrix} = \begin{pmatrix} div\left(I_{d,t,1} * R_1\right) \\ \dots \\ div\left(I_{d,t,L} * R_L\right) \end{pmatrix} \quad (12)$$

Notably, if abnormally large potential value occurs at an abnormal place such as system boundary in rural area where there should not be much people moving, this kind of irregular value must be removed as abnormal value.

**Step 3 (RGM part):**
**Step 3-1.** Determine the boundary of the observation area on the map (shape of the electric circuit network). The shape of the boundary can be arbitrary as long as 1) all the resistors are connected in the network, and 2) the network includes the origin and destination nodes (we have proved that the RGM solution does not rely on the boundary of the system in Supplementary Material 3).

**Step 3-2.** Set a linear equation system $K * I = b$, where $I$ is the current vector to be calculated; $b$ is the vector in which only the last element is the given human voltage 1, and other elements are 0); and $K$ is the coefficient matrix of the linear equation system, which is subject to the following three conditions (Kirchhoff's law). The number of links is denoted as $N$, and $I$ and $b$ are expressed as follows:

$$I = \begin{pmatrix} I_1 \\ I_2 \\ \vdots \\ I_n \end{pmatrix}_{N*1} \quad (13)$$

$$b = \begin{pmatrix} 0 \\ \vdots \\ 0 \\ 1 \end{pmatrix}_{N*1} \quad (14)$$

Condition 1) The divergence of the current of all the nodes in the system is equal to 0, except for the origin and destination:

$$div(I_L) = 0, for\ L \neq origin, destination \quad (15)$$
$$where\ div(I_L) = \sum_{i=1}^{4} I_{(L,\sigma_i)}$$

Zero divergence means the volume flowing in equals to the one flowing out (people will not stay at a node).

Condition 2) The rotation of voltage in the system equals 0:

$$rot(I_L * R_L) = 0, for\ any\ location\ L \quad (16)$$

$$where\ rot(I_L * R_L) = \sum_{k=0}^{3} I_{(\prod_{i=0}^{k} \sigma_i(L), \sigma_{k+1})} * R_{(\prod_{i=0}^{k} \sigma_i(L), \sigma_{k+1})}$$

$$\sigma_0 = 1,\ \sigma_1 = \sigma_{+x},\ \sigma_2 = \sigma_{+y},\ \sigma_3 = \sigma_{-x},\ \sigma_4 = \sigma_{-y}$$

This means the following:

$$rot(I_L * R_L) = I_{(L,\sigma_{+x})} * R_{(L,\sigma_{+x})} + I_{(\sigma_{+x}(L),\sigma_{+y})} * R_{(\sigma_{+x}(L),\sigma_{+y})} -$$

$$I_{(L,\sigma_{+y})} * R_{(L,\sigma_{+y})} - I_{(\sigma_{+y}(L),\sigma_{+x})} * R_{(\sigma_{+y}(L),\sigma_{+x})}$$

Zero rotation means people will not be wandering around in the same area all the time.

Condition 3) Arbitrarily find one path from the origin to destination such that the summation of the voltage on the path equals 1:

$$\sum_{i \in P}(I_i * R_i) = 1 \quad (17)$$

where link $i$ is on any path $P$ from the origin to destination

As a simple case, we selected a rectangular observation area with m rows and n column nodes in the system (boundary as shown in Fig. 3(a)). Overall, there were $N = (m-1)*n + (n-1)*m$ links, $M_1 = m*n$ nodes, and $M_2 = (m-1)*(n-1)$ rotations in the system. These three conditions can be summarized as follows (we set $\sigma_{+x}$ and $\sigma_{+y}$ as the positive direction, and $\sigma_{-x}$ or $\sigma_{-y}$ as the negative direction):

$$K = \begin{pmatrix} K_{1,(M_1-2)*(N)} \\ K_{2,(M_2)*(N)} \\ K_{3,(1)*(N)} \end{pmatrix}_{N*N} \quad (18)$$

where $(K_1)_{i,j} = \begin{cases} 1, \text{if link } j \text{ is connected with node } i \text{ with direction } \sigma_{+x} \text{ or } \sigma_{+y} \\ -1, \text{if link } j \text{ is connected with node } i \text{ with direction } \sigma_{-x} \text{ or } \sigma_{-y} \\ 0, else \end{cases}$

$(K_2)_{i,j} = \begin{cases} R_j, \text{if link } j \text{ is in the rotation of the node } i \text{ with direction } \sigma_{+x} \text{ or } \sigma_{+y} \\ -R_j, \text{if link } j \text{ is in the rotation of the node } i \text{ with direction } \sigma_{-x} \text{ or } \sigma_{-y} \\ 0, else \end{cases}$

$(K_3)_{i,j} = \begin{cases} R_j, \text{if link } j \text{ is on the path } P \text{ with direction } \sigma_{+x} \text{ or } \sigma_{+y} \\ -R_j, \text{if link } j \text{ is on the path } P \text{ with direction } \sigma_{-x} \text{ or } \sigma_{-y} \\ 0, else \end{cases}$

Since $(M_1 - 2) + (M_2) + (1) = N$, and each condition is independent[48], the coefficient matrix is full-rank, meaning that there is a unique solution to the linear equation system according to linear algebra.

**Step 3-3**. Learn the human resistance value from the real data by using the method in RECM, solve the linear equations in step 3-2 to generate the human current, and plot it on the map to observe the manner in which people flow from the origin to destination.

**Step 3-4 (Optional)**. If humans cannot pass through some node for some reason, set up an "accident-affected area" in the map, that is, set the value of resistance on the link around the accident-affected area to infinity (a large number in numerical calculation) to simulate the human flow when accident area occurs.

## 4.5 Time complexity of the algorithms

Regarding to time complexity of the algorithms, the time complexity in calculating the resistance was $O(iter * d * t * L)$ in the previous study, whereas that in our study was $O(d * t * L)$ after our improvement. In Tokyo, the number of nodes, iterations, days, and time periods were approximately $10^4$, $10^4$, $10^2$, and $10^1$, respectively. Therefore, the time cost of the number of nodes and iterations was dominant, and the computational cost was reduced to 1/10000 of the original cost. In a real-world scenario, it took approximately 1–2 h to calculate the current, 3–5 days (using the Adam algorithm[49] before, but now only 1 h) to calculate the resistance, and 1–2 hours to calculate the electric potential for the Greater Tokyo Area, which had approximately 30000 nodes (node size: 500 * 500 m; computational environment shown in Method). Using the RECM, the same result was obtained at a computational cost that was drastically decreased, indicating that we made it easier to deploy and more useful for other researchers. To use the RGM, a linear equation system needs to be solved. Its theoretical time complexity is $O(n^3)$, where $n$ is the number of nodes in the system. Since its coefficient matrix is sparse, the method proposed by Peng et al.[50] for solving the linear equation can be utilized to practically reduce the time complexity cost to $O(n^{2.33})$.

**Data Availability.** Data used in this study can be purchased from a Japanese private company, Agoop, which sells "The location information big data which acquired from the smart phone app". The product name in Japanese is "Pointo-gata ryudou-jinkou data" ("Point-type population data").

**Author Contributions.** M.T. was the leader of this project, designed the whole research plan and directed writing of the manuscript. Z.Z. analyzed the raw data, performed the numerical calculation, maintain open-source code, and wrote the manuscript. H.T. and Z.Z. developed methods of data analysis, M.T. and H.T. revised the manuscript.

**Additional information.** The open-source codes in this study can be downloaded from the author's Github https://github.com/Zhihua-Zhong/Revised-Electric-Circuit-Model.

**Competing interests statement.** The author(s) declare no competing interests.

**Funding.** This work was partially supported by the "Research Project for Overcoming Coronavirus Disasters" by the Institute of Innovative Research, Tokyo Institute of Technology; Grant-in-Aid for Scientific Research (B) (Japan Society for the Promotion of Science; Grant Number 18H01656 and 22H01711); and "Support for Pioneering Research Initiated by the Next Generation" by Japan Science and Technology Agency (Grant Number JPMJSP2106).

**Acknowledgement.** We thank Dr. Jun-ichi Ozaki and Dr. Yohei Shida for their valuable discussions and Mr. Hajime Koike for suggesting a useful method of mapping the map of Japan into relative coordinates $L = (x, y)$ in our study. We thank Agoop for providing GPS datasets and Editage (http://www.editage.com) for English language editing.

algorithms (SODA), pp. 504-521. Society for Industrial and Applied Mathematics, 2021.

**Figure legends**

Figure 1. Human current. Fig. 1 (a–e): Illustration of the working of the model. Fig. 1 (f-h): Time series of the population, mean velocity, and human current (blue, orange, and green lines represent Tokyo Station with the high population in the city center, Nagatsuda Station with the medium population, and Kazo with the low population in the rural area from Sep 5 to Sep 9, 2022), respectively. Fig. 1 (i): Human current cumulative distribution with respect to different directions in semi-log plot. Fig. 1 (i-1, i-2, and i-3): The red, orange, and blue lines represent the distributions of the absolute, positive (direction right), and negative (direction left) values of human current, respectively. Fig. 1 (i-4): The blue and orange lines represent the distribution of the absolute value of human current in the right and up directions, respectively. The guide lines show the average slopes when approximated by exponential distributions.

Figure 2. Revised electric circuit model (RECM). Fig. 2 (a-b): Spatial distribution of the human conductivity in the greater Tokyo area. Fig. 2 (c): Human voltage time series from Sep 5 to Sep 9, 2022 (the blue, orange, and green lines represent Tokyo Station with the high population in the city center, Nagatsuda Station with the medium population, and Kazo with the low population in rural area, respectively). Fig. 2 (d): Spatial distribution of electrical energy dissipation which characterizes highly congested currents (time period is from 7:30 a.m. to 8:00 a.m. in the morning rush hour). Fig. 2 (e–g): Human potential patterns at morning, noon, and evening. Fig. 2 (h): Human potential time series on Sep 1, 2022.

Figure 3. Comparison between different cities. Fig. 3 (a): Bar plot of population and maximum conductivity in each city (population data is collected from: Tokyo Metropolitan Government. 2023. https://www.metro.tokyo.lg.jp/english/index.html). Fig. 3 (b). Linear regression analysis on population and conductivity. Fig. 3 (c-d): CDF of human resistance and observed potentials (in morning rush hour from 7:30 am to 8:00 am) in each city. Fig. 3 (e): The maximum and minimum potential values for different periods in each city. Fig. 3 (f): Linear regression analysis on the average radius of each city and the range of potential values (calculated in the morning rush hour from 7:30 am to 8:00 am).

Figure 4. Route generation model. Fig. 4 (a): Routes generated by the RGM (the origin and destination are Ayase Station and Shinanomachi Station, respectively, which are typical stations located at north-east and south-west area of Tokyo, respectively). Fig. 4 (b): Routes recommended by Google Maps (Google Map. Recommendation route from Ayase station to Shinanomachi station. 2023. https://www.google.com/maps/dir/Ayase+Sta.,+3+Chome-1+Ayase,+Adachi+City,+Tokyo/Shinanomachi,+Shinjuku+City,+Tokyo/@35.7000431,139.7195152,13z/data=!4m14!4m13!1m5!1m1!1s0x60188fb9cc8beb47:0x48450c07579a074!2m2!1d139.8254234!2d35.7622297!1m5!1m1!1s0x60188c92379c0951:0x907dd02bc4813974!2m2!1d139.7198607!2d35.6817841!3e3?entry=ttu). Fig. 4 (c): Routes generated by the RGM when the accident-affected area is set up (when a natural disaster occurs, and the railway or highway cannot function normally, it can be viewed as an accident-affected area, which is represented by the black rectangle in the figure; in this case, the resistance value near the accident-affected area will be set to infinity, which is a large number)

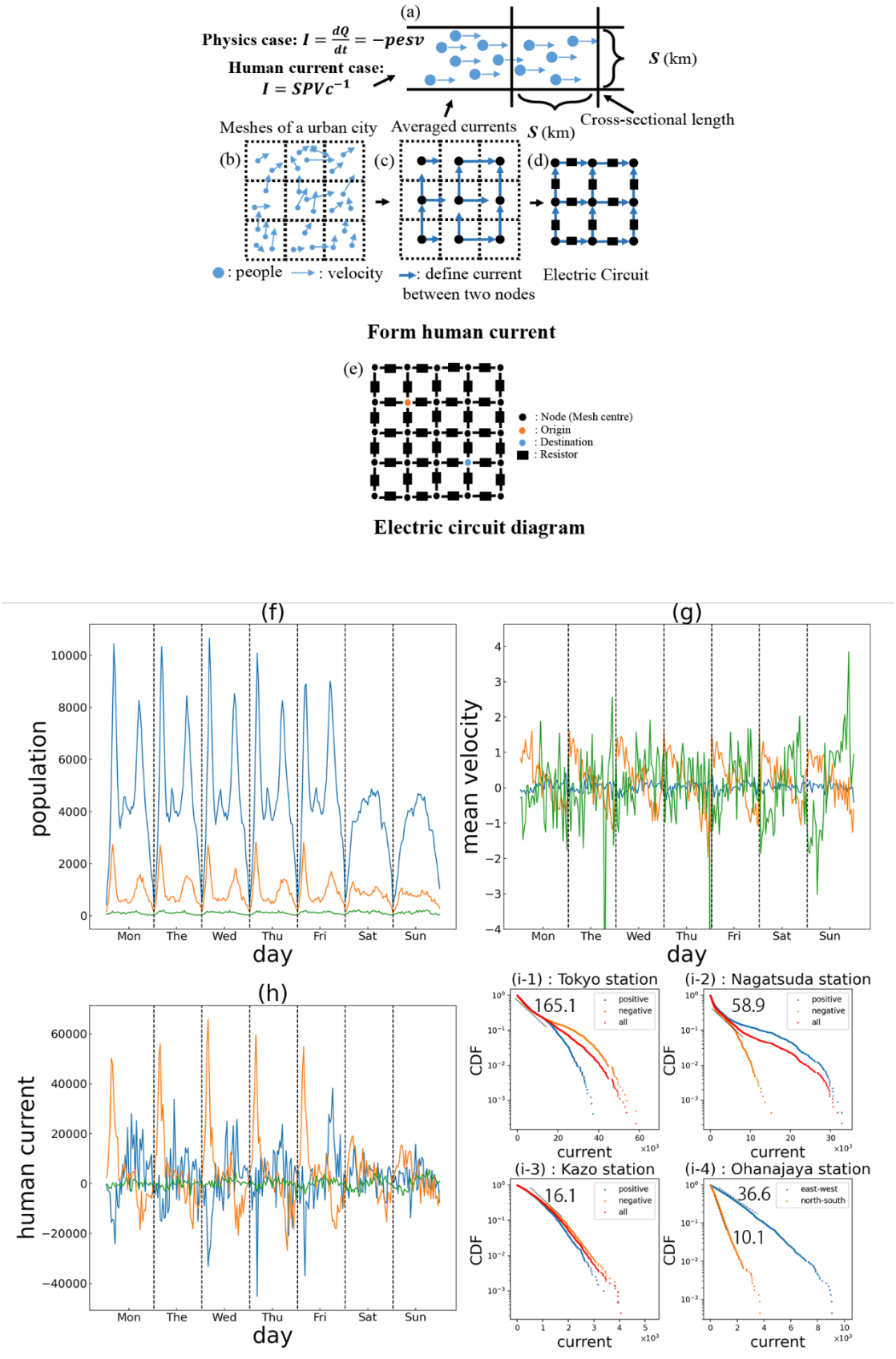

Figure 1. Human current

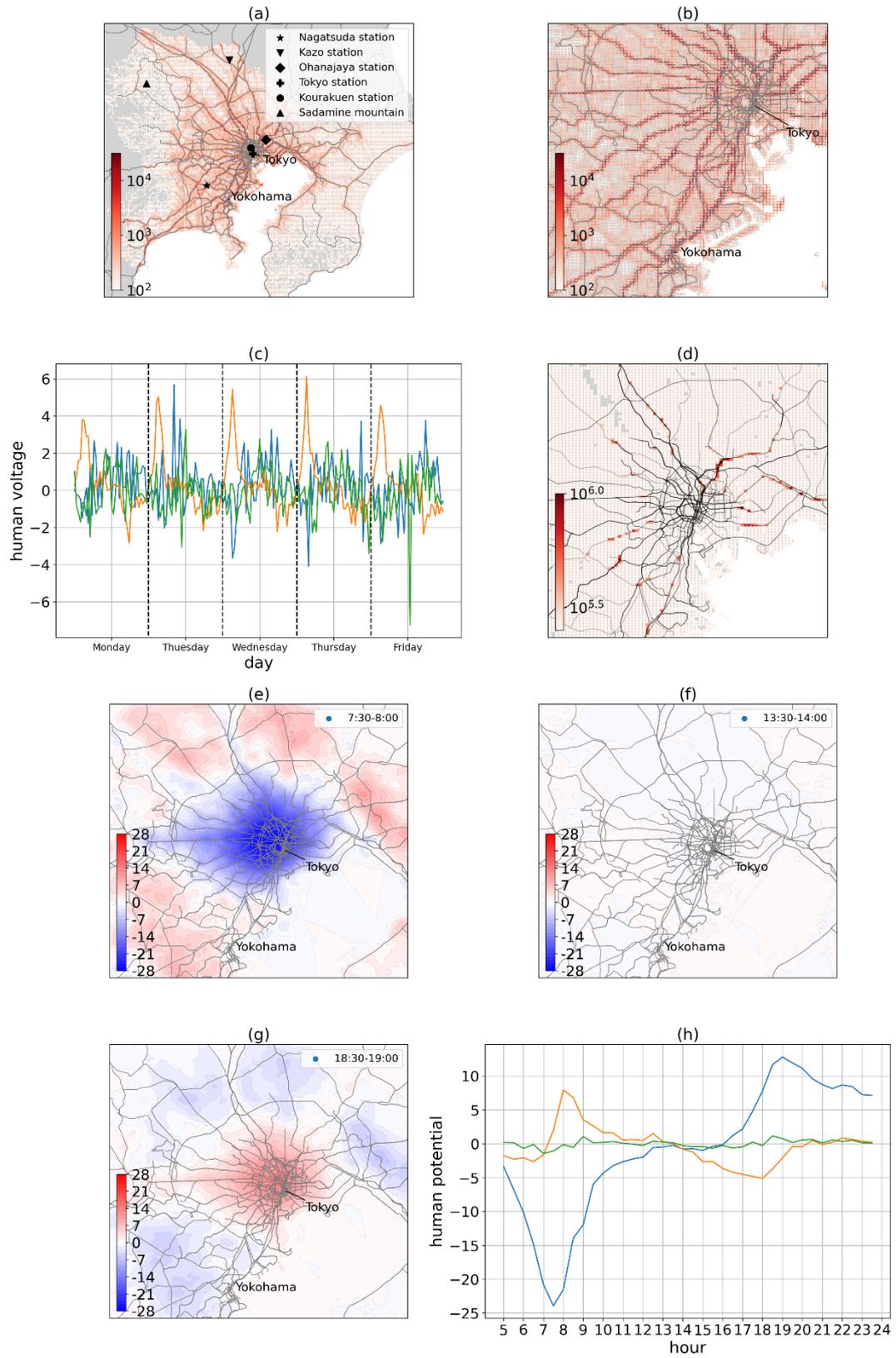

Figure 2. Revised Electric Circuit Model

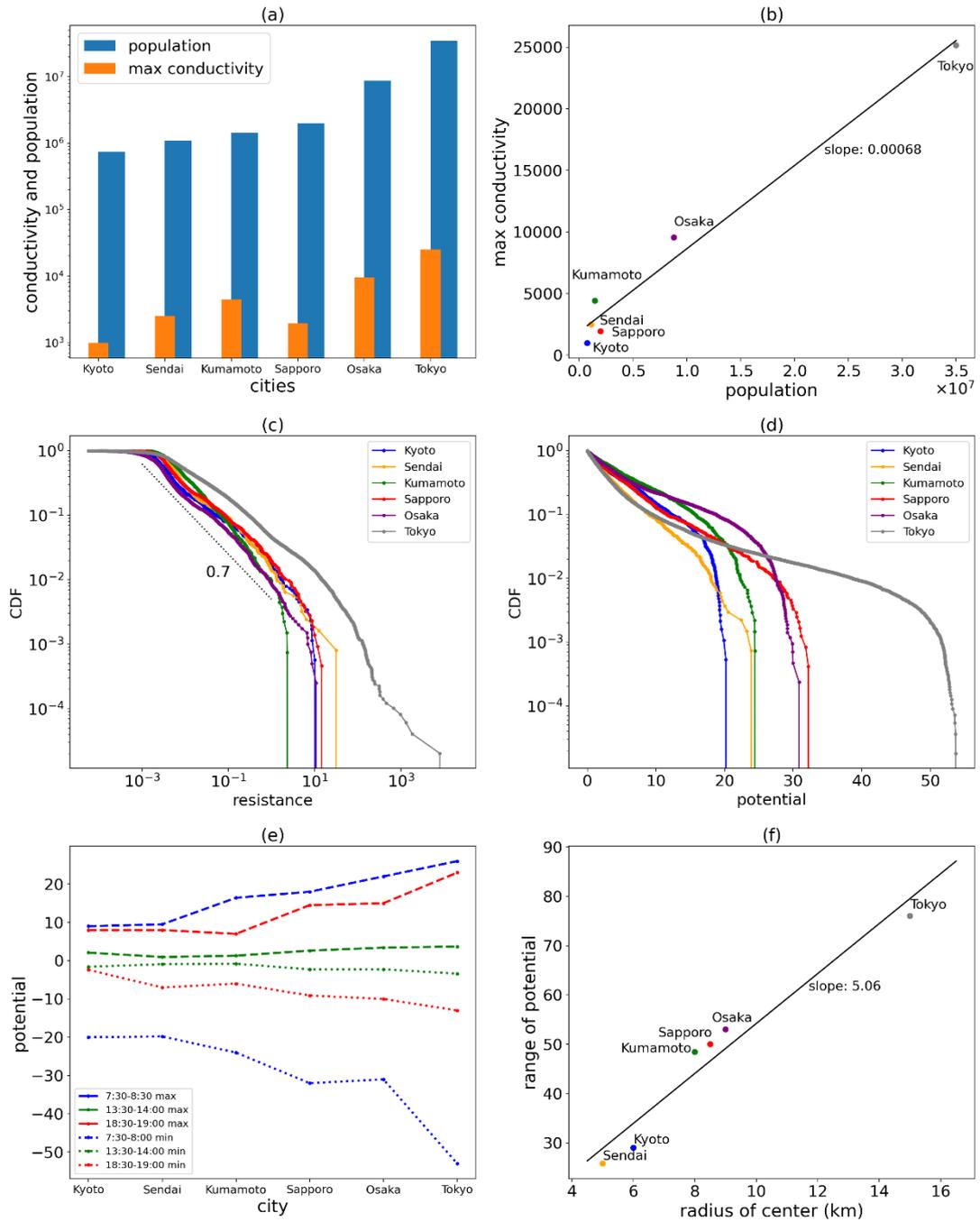

Figure 3. Comparison between different cities

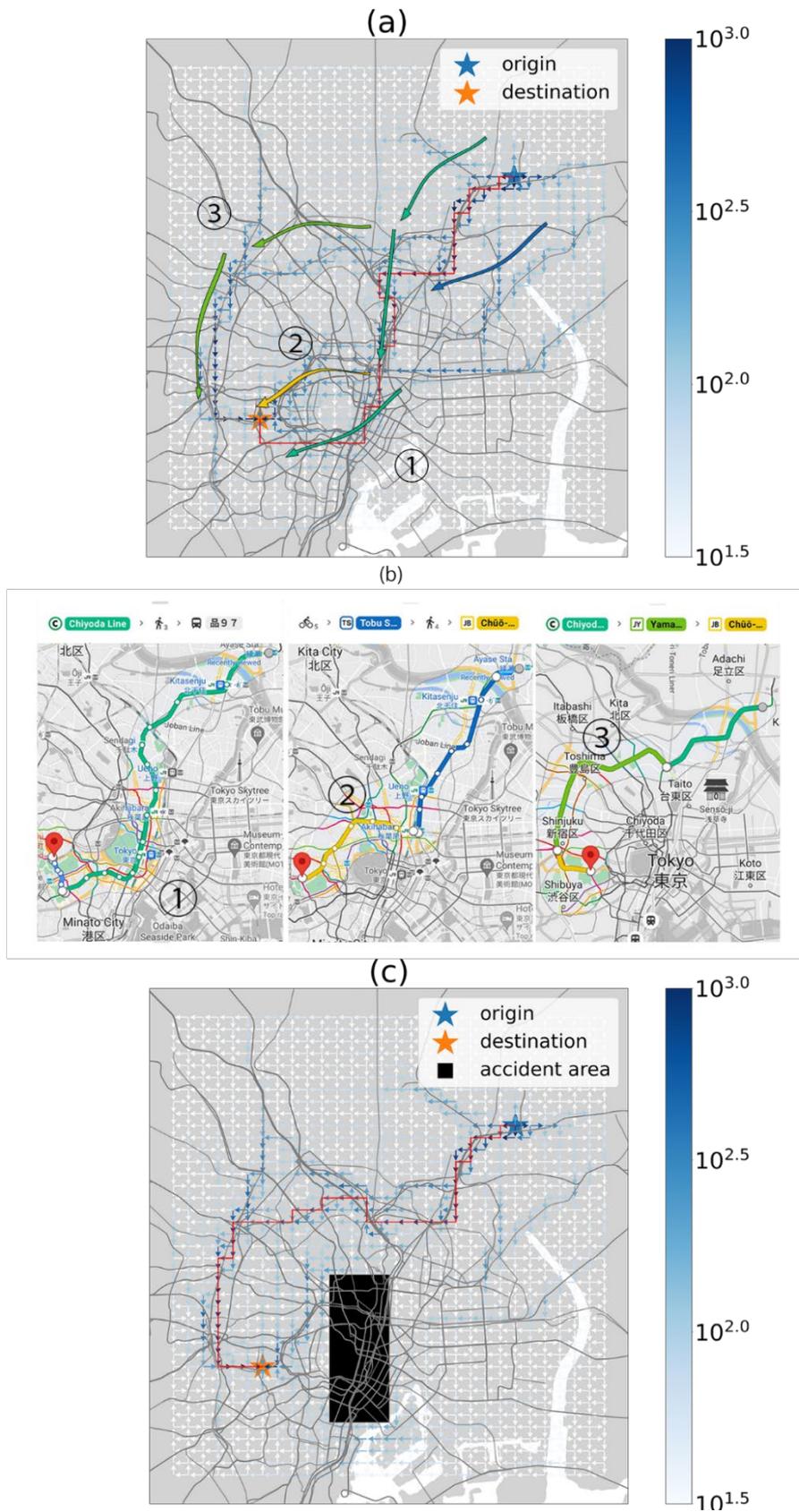

Figure 4. Route Generation Model

**Novel approaches to urban science problems: human mobility description by physical analogy of electric circuit network based on GPS data**

**Supplementary Material**

**Authors:** Zhihua Zhong[1], Hideki Tayakasu[1,2], Misako Takayasu[1]

**Institutions and Affiliations:**

[1] *School of Computing, Tokyo Institute of Technology, Tokyo, Japan.*

[2] *Sony Computer Science Laboratories, Tokyo, Japan.*

**Correspondence Author:**

Misako Takayasu

School of Computing, Tokyo Institute of Technology, 2-12-1 Ookayama, Meguro-ku, Tokyo, Japan, 152-8550

Email: takayasu.m.aa@m.titech.ac.jp

# Contents

## 1. Brief introduction

In supplementary, detailed content which could not be included in the main part due to words limitation will be discussed. In section 2, we will discuss the difference between new (in our research) and old (in previous research[1]) resistance and rotation, showing that even we changed the definition of resistance, results in the previous research still can be reproduced. In section 3.1, we recommend our new method (using the mean value of current to calculate resistance) and explain its three advantages. We show that the solution of RGM (route generation model) does not rely on the system's boundary, and people can choose the observation area they are interested in by generating a route between the origin and destination in 3.2. More examples of the route generated by RGM will be shown in 3.3. Moreover, the detailed method of calculating temporal human potential was not written clearly in the previous paper, and we will complement it as a supplement to the former research in section 4. Finally, the place our open-source code release and the tutorial to use our code will be introduced in section 5.

**Raw GPS data examples:**

Examples of raw GPS data are shown in Fig. 1. Researchers with GPS data including time, longitude, latitude, speed, and course (angle) can conduct research about human flow pattern by using our model.

|  | dailyid | hour | minute | longitude | latitude | speed | course |
|---|---|---|---|---|---|---|---|
| 0 | b6db370c5ba96648d0599cb626f48cd28c6a247705b837... | 0 | 0 | 139.752251 | 35.648805 | NaN | 276.674 |
| 1 | b1bf35c365b7ca77c8b50b3098030bf5f533a1c53a4bd3... | 0 | 0 | 139.664415 | 35.502598 | NaN | 228.814 |
| 2 | 7a1592328c65d22f6e23055572c6b1bdc27b689b06833c... | 0 | 0 | 139.767463 | 35.681241 | NaN | 273.212 |
| 3 | f5141e0f14ff8a8888336e59d02d9c78edb1be3500f09f... | 0 | 0 | 139.775246 | 35.724509 | NaN | 164.375 |
| 4 | 162b957019df9f8731eefe68022f8ed9b47aa18fcb786a... | 0 | 0 | 139.434758 | 35.873877 | 0.000 | 316.398 |
| ... | ... | ... | ... | ... | ... | ... | ... |
| 95967702 | 68c59462a32d565934f3597209304ef979a668613774fe... | 23 | 59 | 139.704592 | 35.690517 | 1.280 | 27.470 |
| 95967703 | 4a2b3c8031a526742567345ab92fec20fd66f946b2a9d0... | 23 | 59 | 139.618750 | 35.479167 | 0.000 | NaN |
| 95967704 | 47fbb67bc2c5a9ea8506e9ec4c8c0bd65cfa8faa291d04... | 23 | 59 | 139.731250 | 35.804167 | 0.000 | NaN |
| 95967705 | 1e4bd6b2153293a044f384e8f06995212fa6546af9ed74... | 23 | 59 | 139.668750 | 35.670833 | 2.796 | 336.410 |
| 95967706 | 431d88aeeeca05f0e99ae3582708b1d1a22a7b8a7447bf... | 23 | 59 | 139.906250 | 35.679167 | 0.539 | 140.860 |

Figure 1. example of raw GPS data (Note: speed and course are not necessary for raw data because it can be calculated if (user, time, longitude, latitude) are known)

## 2. Discussion about new and old resistance and rotation

In order to show that our new method can replace the old method to calculate resistance in the previous research, in this section, the property of resistance will be discussed in detail. We will introduce the method to calculate resistance in the previous research in 2.1, explain four limitations of the method proposed in the previous research in 2.2, describe our idea of new methods to calculate resistance in detail in 2.3, show that human current is subject to exponential distribution in 2.4, and compare the new and old resistance and rotation in 2.5.

**2.1 Review of the method to calculate resistance in the previous research**

In the previous research[1], the authors calculated resistance by the following inspiration: in physics, there should be no rotation in electric field of an electric circuit with stationary current. They defined the rotation at day $d$, time $t$, and location $L$ by the following formula:

$$rot(I_{d,t,L} * R_L) = I_{(d,t,L,\sigma_{+x})} * R_{(L,\sigma_{+x})} + I_{(d,t,\sigma_{+x}(L),\sigma_{+y})} * R_{(\sigma_{+x}(L),\sigma_{+y})} - I_{(d,t,L,\sigma_{+y})} * R_{(L,\sigma_{+y})} - I_{(d,t,\sigma_{+y}(L),\sigma_{+x})} * R_{(\sigma_{+y}(L),\sigma_{+x})}$$

For simplicity in notation this equation is written as:

$$rot(I_{d,t,L} * R_L) = \sum_{k=0}^{3} I_{(d,t,\prod_{i=0}^{k} \sigma_i(L),\sigma_{i+1})} * R_{(\prod_{i=0}^{k} \sigma_i(L),\sigma_{i+1})} \tag{1}$$

where, $\sigma_0 = 1: (x,y) \to (x,y), \sigma_1 = \sigma_{+x}: (x,y) \to (x+1,y), \sigma_2 = \sigma_{+y}: (x,y) \to (x, y+1), \sigma_3 = \sigma_{-x}: (x,y) \to (x-1,y), \sigma_4 = \sigma_{-y}: (x,y) \to (x, y-1)$

They calculated the resistance value by treating it as an optimization problem whose objective function is the square of the summation of the rotation of nodes around the map, as shown in the following formula. The optimization problem is subject to, firstly that $R_{(l,\sigma)}$, resistance value at location $L$ and direction $\sigma$, should be greater than 0, and secondly they assume a kind of conservation rule that $\sum_{L,\sigma} R_{(L,\sigma)}^{(0)} = \sum_{L,\sigma} R_{(L,\sigma)}^{(\tau)}$ ($\tau^{th}$ iteration's summation of resistance equal to $1^{st}$ iteration's summation of resistance).

$$\text{minimize } L = \sum_{d,t,L} \left(rot(I_{d,t,L} * R_L)\right)^2 - \gamma * \sum_{(L,\sigma)} Log(R_{(L,\sigma)}) \tag{2}$$

$$\text{subject to: } \sum_{L,\sigma} R_{(L,\sigma)}^{(0)} = \sum_{L,\sigma} R_{(L,\sigma)}^{(\tau)}$$

where, $R_{(L,\sigma)}^{(0)} = 1$ for each location $L$ and direction $\sigma$

Adam[4], an optimization algorithm popular in deep learning, was employed to solve this optimization problem to get the values of resistance. Adam will update resistance by the following form:

$$R_{(L,\sigma)}^{(\tau)} = R_{(L,\sigma)}^{(\tau-1)} - \alpha * f(\frac{\partial L}{\partial R_{(L,\sigma)}^{(\tau-1)}}) \tag{3}$$

where $\alpha$ is learning rate

$$\frac{\partial L}{\partial R_L} = (2 * I_{d,t,L} * \Sigma_{d,t} rot(I_{d,t,L} * R_L) - \gamma * \frac{1}{R_L} \quad (4)$$

There are cases, $R_{(L,\sigma)}^{(\tau)}$ drops below 0 after an update, if so, authors adjusted the parameter $\gamma$ to $10 * \gamma$, $10^2 * \gamma$, $10^3 * \gamma$, …, until the value of $R_{(L,\sigma)}^{(\tau)}$ becomes positive once again. Meanwhile, to keep $\Sigma_{L,\sigma} R_{(L,\sigma)}^{(0)} = \Sigma_{L,\sigma} R_{(L,\sigma)}^{(\tau)}$, authors applied a normalization after each iteration by the following formula (5):

$$R_{(L,\sigma)}^{(\tau)} := R_{(L,\sigma)}^{(\tau)} * (\frac{\Sigma_{L,\sigma} R_{(L,\sigma)}^{(\tau-1)}}{\Sigma_{L,\sigma} R_{(L,\sigma)}^{(\tau)}}) \quad (5)$$

## 2.2 Limitation of the old method

There are mainly four limitations to calculating resistance using the method suggested in the previous research.

The first problem is the high computational cost of calculating resistance. Our computational environment is CPU: 2 * Intel Xeon Gold 6248R (24 core); Memory: 500 GB; OS: Linux Ubuntu. In real-time, it took about 1-2 hours to calculate current, 3-5 days to calculate resistance, and 1-2 hours to calculate the electric potential for the greater Tokyo area, with about 30000 nodes (node size: 500m * 500m) in the real world time. The computational cost problem of human resistance need to be solved, otherwise the application of ECM will be limited.

The second problem is the difficulty in fine-tuning the hyperparameter of Adam. When using ECM on a new city's data, there is no standard way to find the best parameter, therefore trying different combinations of parameters and waiting a long time before checking whether it works is the only thing can do, which means the old model is not easy to be deployed and used.

The third problem is that the calculated value of resistance depends on the initial value and hyperparameter of Adam. Even though people manage to make Adam converge, resistance value will not be unique. Suppose researchers have two cities' data, such as Tokyo's and New York's. In that case, their suitable hyperparameter may differ, and directly comparing the value of resistance calculated between two cities is meaningless. Moreover, because the electric potential calculation is based on the value of resistance, comparing the potential value between different cities is also meaningless. Above problems can be solved by our proposed new method. Therefore, it is interesting to conduct further research, like comparing the difference of human electric potential between different cities.

The fourth problem is that the method in the previous research will cause some outliers (calculation errors) when calculating resistance for rural areas. Fig. 2 shows the histogram of resistance calculated by our method and the old method on the left and the right, respectively. There is a peak on the far right of the histogram of old resistance. It is because the current value is very small in rural areas, which account for 70% of the greater Tokyo area. In Eq. (3-4), when the current is small, $\frac{\partial L}{\partial R_L}$ is also tiny, making $R_{(L,\sigma)}^{(\tau)}$ cannot update, and the value of $R_{(L,\sigma)}^{(\tau)}$ always equals the initial value of 1. However, in Eq.(5), the normalization term value $\frac{\sum_{L,\sigma} R_{(L,\sigma)}^{(\tau-1)}}{\sum_{L,\sigma} R_{(L,\sigma)}^{(\tau)}}$ is slightly greater than one because after each time updated, the resistance value $\sum_{L,\sigma} R_{(L,\sigma)}^{(\tau)}$ will become smaller. Therefore, after 10,000 iterations, resistance in the rural area is approximately equal to $1*\left(\frac{\sum_{L,\sigma} R_{(L,\sigma)}^{(\tau-1)}}{\sum_{L,\sigma} R_{(L,\sigma)}^{(\tau)}}\right)^{10000} \approx 2-3$. These rural areas occupy a large amount of ratio in Tokyo, and it is the reason for the appearance of the peak (calculation error) on the right of Fig. 2. However, the calculation of human potential mainly relies on the current value in the nodes near the city centre, not the rural area. Therefore, human potential can still be correctly calculated in the previous research.

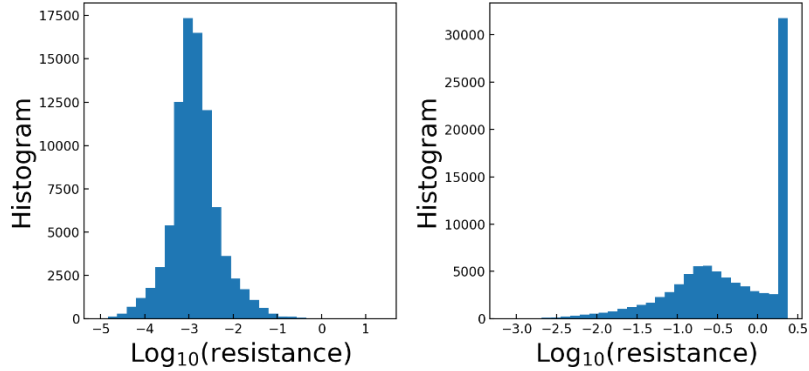

Figure 2. Histogram of new and old resistance

(left: method in our research, right: method in previous research)

**2.3 New methods to calculate the resistance**

To improve ECM, we paid attention to a relation reported in the previous research, that human conductivity, as the inverse of resistance $\rho_{(L,\sigma)}$, at location $L$, direction $\sigma$ is proportional to the maximum absolute current value $\max_{d,t}\{|I_{(d,t,L,\sigma)}|\}$ at location $L$, direction $\sigma$. That is:

$$\rho_{(L,\sigma)} \propto \max_{d,t}\{|I_{(d,t,L,\sigma)}|\} \tag{6}$$

Because conductivity is the inverse of resistance, resistance value equals a constant k divided by the maximum absolute current value. Let's assume simply that constant k equals 1. Then resistance can be calculated by the inverse of its maximum current. (Method 1)

$$R_{(L,\sigma)} = \frac{1}{max_{d,t}\{|I_{(d,t,L,\sigma)}|\}} \quad (7)$$

Based on the resistance calculated by Method 1, temporal human potential can be calculated and compared with the one calculated in previous research. Fig. 4 shows a robust linear relationship between new and old results. R square of linear regression is 0.95, which means people can use the potential value calculated by our new method to replace the old one.

We tried to plot the CDF (cumulative density function) of absolute value current in many famous locations in Tokyo, such as Kourakuen, Kazo, Hakone, and Tokyo station as shown in Fig. 3. CDF shows that the current distribution in these nodes approximated by the exponential distribution. However, fluctuation of the maximum value is generally very large, so using the maximum current to determine resistance may not be robust.

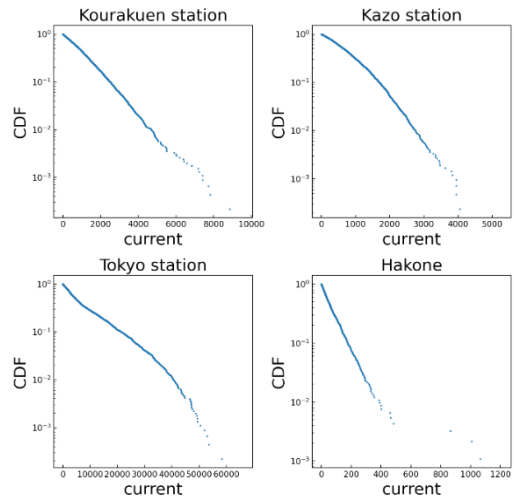

Figure 3. CDF of current in different places

In order to find a better definition, we searched the cases of 97.5% quantile (Method 2), 95% quantile current (Method 3), and the parameter of exponential distribution of current (the mean absolute current, Method 4) to determine resistance.

$$R_{(L,\sigma)} = \frac{1}{Quantile_{0.975,d,t}\{|I_{(d,t,L,\sigma)}|\}} \quad (8)$$

$$R_{(L,\sigma)} = \frac{1}{Quantile_{0.95,d,t}\{|I_{(d,t,L,\sigma)}|\}} \quad (9)$$

Fig. 4 shows the result of the relationship between old potential and new potential calculated by different methods. When using 97.5% quantile and 95% quantile current to calculate resistance,

R square of linear regression improved, which confirms our conjecture that maximum current is not robust.

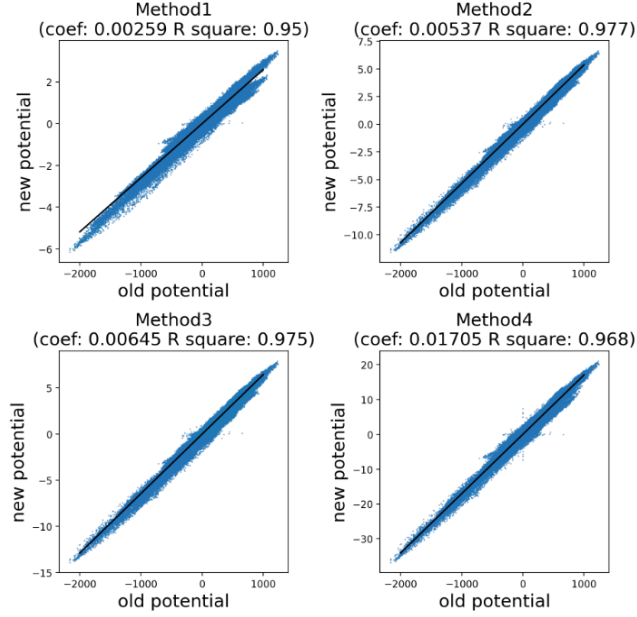

Figure 4. Comparison of potential calculated by the new and old method
(Note: method 1: maximum current, method 2: 97.5% quantile current,
method 3: 95% quantile current, method 4: mean current)

Time series for a different day of different per cent quantile current on a certain day was plotted in Tokyo Station (left: area with high population) and Hakone (right: area with low population), as shown in Fig. 5. We denote current time series at a specific quantile $q$, at day $d$, at location $L$ as:

$$CTS_{(q,d,L)} = Quantile_{q,t}\{|I_{(d,t,L)}|\} \tag{10}$$

In Fig. 5, we observe that the current series $CTS_{q,d,L=Tokyo\ Station}$ fluctuate strongly when quantile $q$ is high ($q \in [0,1]$). For example, we pay attention to the blue line (maximum current of every day) in Tokyo station, $CTS_{(q=1,d,L=Tokyo\ station)}$. If the observation period is $d_1 = [1, 10]$ or $d_2 = [11, 20]$, the difference between the value of the right current (blue line) gained in $d_1$ and $d_2$ is large (12500 vs 17500). The resistance value is expected to be independ of the observation period because resistance should be a quantity reflecting the infrastructure level of a place independent of time. Therefore, it is better to use a per cent quantile $q$ that the standard deviation of $CTS_{q,d,L}$ is small, which means that no matter which time people observe, the resistance value will not change too much.

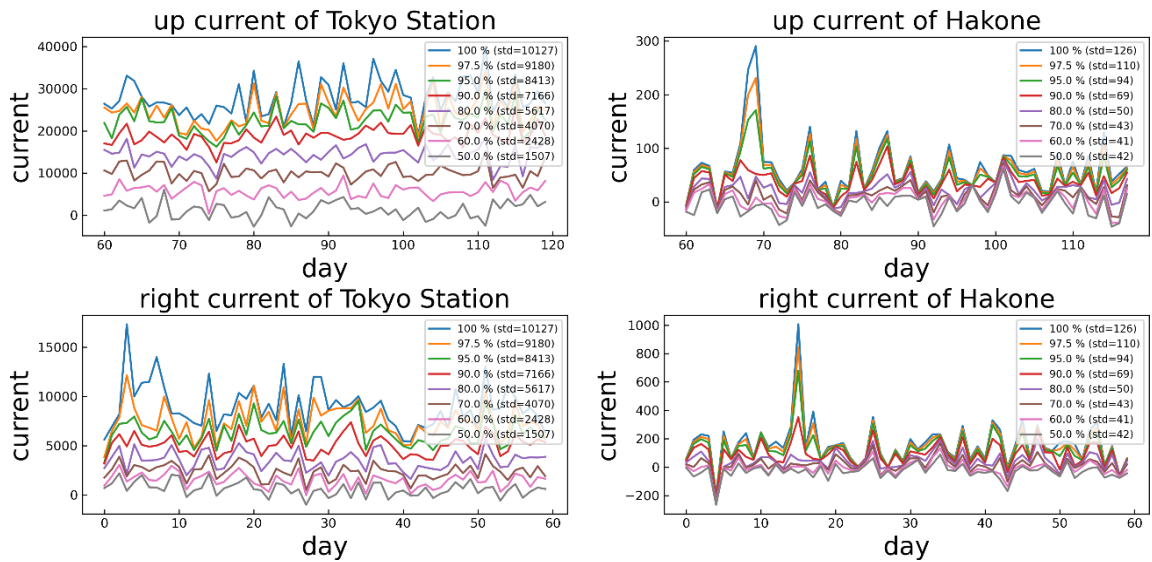

Figure 5. Time series of current at different places (left: Tokyo Station, right: Hakone)

Relation between the quantile and standard deviation of $CTS_{q,d,L}$ was plotted in Fig. 6. When the quantile decreased, the standard deviation dropped rapidly at the beginning and became steadier later, which inspired us to use Elbow Method to determine the suitable quantile. For example, the shape of the curve looks like the human arm, before 95%, standard deviation changes rapidly, and after that, it gradually converges. Therefore, we can use a percentage of 95% or 92.5% to determine resistance.

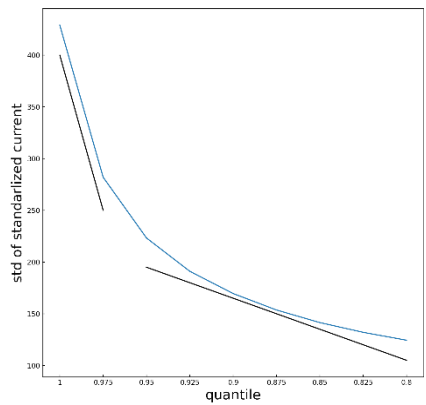

Figure 6. Relationship between quantile and standard deviation of $CTS_{q,d,L}$

## 2.4 Current is subject to exponential distribution

As we mentioned before, in Fig. 3, we noticed a phenomenon: current distributions in many places are close to the exponential distribution. Therefore, K-S test was employed to check whether the current is subject to exponential distribution in most places in the greater Tokyo area. In Fig. 7, areas where the p-value of the K-S test is greater than 0.05, are shown in blue, which means that in most places, human current distribution can be treated as exponential distribution.

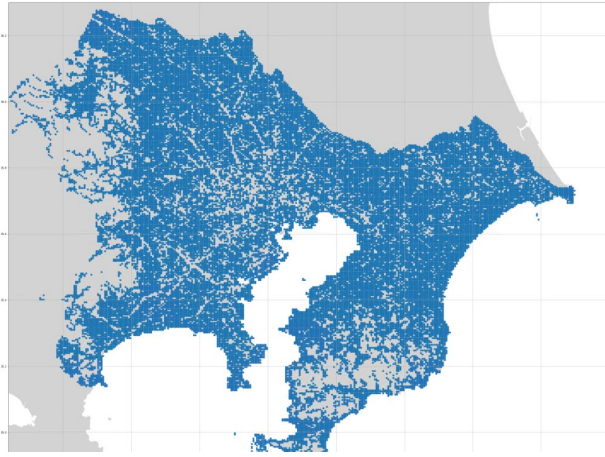

Figure 7. KS-test result in the greater Tokyo area (blue area means: p-value is greater than 0.05)

Therefore, we write the PDF (Probability density function) of the current for each location $L$ and direction $\sigma$ in the following form:

$$p(I_{(L,\sigma)}) \sim \left(\frac{1}{I^0_{(L,\sigma)}}\right) * e^{-\frac{|I_{(d,t,L,\sigma)}|}{I^0_{(L,\sigma)}}} \qquad (11)$$

where $I^0_{(L,\sigma)} = \underset{d,t}{mean}\{|I_{(d,t,L,\sigma)}|\}$

Being the parameter of the exponential distribution, $I^0_{L,\sigma}$ means the mean current at location $L$, direction $\sigma$. It gave us a hint that we may also use $I^0_{L,\sigma}$ to determine resistance as follows (Method 4):

$$R_{(L,\sigma)} = \frac{1}{I^0_{(L,\sigma)}} = \frac{1}{\underset{d,t}{mean}\{|I_{(d,t,L,\sigma)}|\}} \qquad (12)$$

As shown in Fig. 4, there is little difference between new and old potential concerning the different proposed methods. Furthermore, we will show there are not much difference in resistance and rotation values no matter which new methods are used in the following section 2.5 by comparing new and old resistance and rotation.

**2.5 Comparison of new and old resistance and rotation**

Fig. (8-9) shows the CDF of resistance and conductivity (infrastructure level) calculated by different methods. When quantile $q$ decrease, the value of the current time series $CTS_{(q,d,L)}$ will decrease, the resistance value increases and the conductivity value decreases.

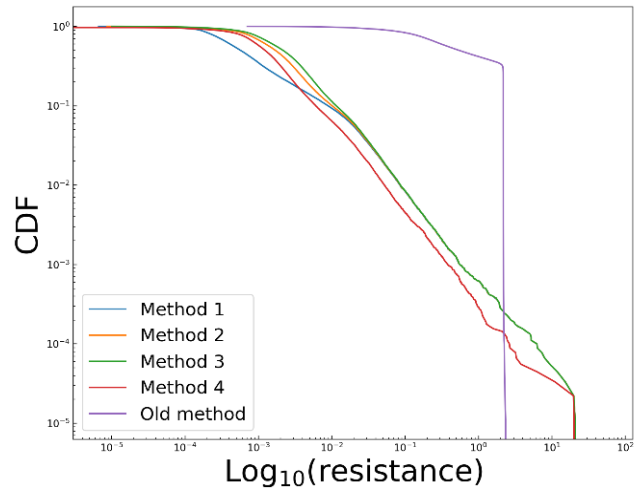

Figure 8. Comparison of CDF of resistance calculated in different methods
(Note: method 1: maximum current, method 2: 97.5% quantile current,
method 3: 95% quantile current, method 4: mean current)

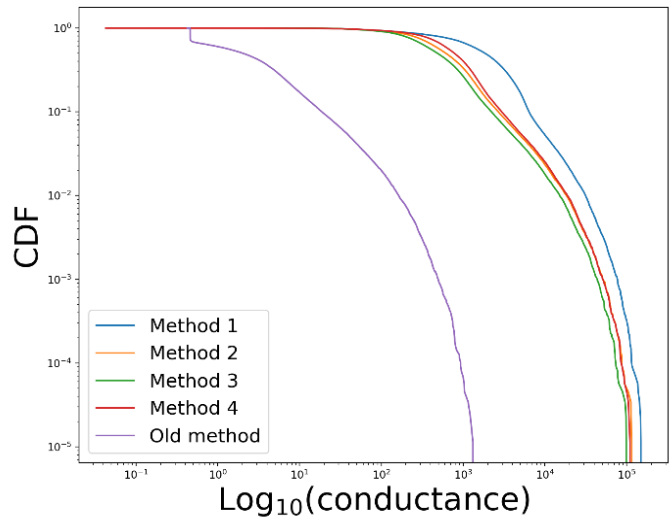

Figure 9. Comparison of CDF of conductivity calculated in different methods
(conductivity is the inverse number of resistance, meaning the infrastructure level)

Fig. 10 shows a linear relationship between new and old resistance. The fluctuation on the far right is caused by the outlier (calculation error) we mentioned in 2.2, Fig. 2. The calculation error problem has also been solved in our new methods.

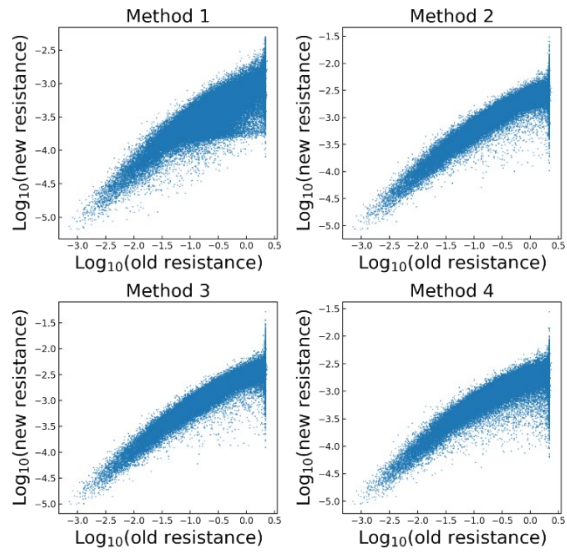

Figure 10. The relationship between old and new resistance concerning different methods
(Note: method 1: maximum current, method 2: 97.5% quantile current,
method 3: 95% quantile current, method 4: mean current)

In the previous research, authors tried to reduce rotation of electric field in the electric circuit, and authors calculated resistance by treating it as an optimization problem minimizing the sum of squared rotation in the former method. While, our new method does not need to consider rotation when calculating resistance. Therefore, a problem naturally emerged regarding the rotation in our case. We calculated rotation and compared the value of the new rotation with the old one. Interestingly, Fig.11 shows a linear relationship between the new and old rotation, which shows that the small value in the old rotation is also small in the new one, and the large value in the old rotation is also large in the new one.

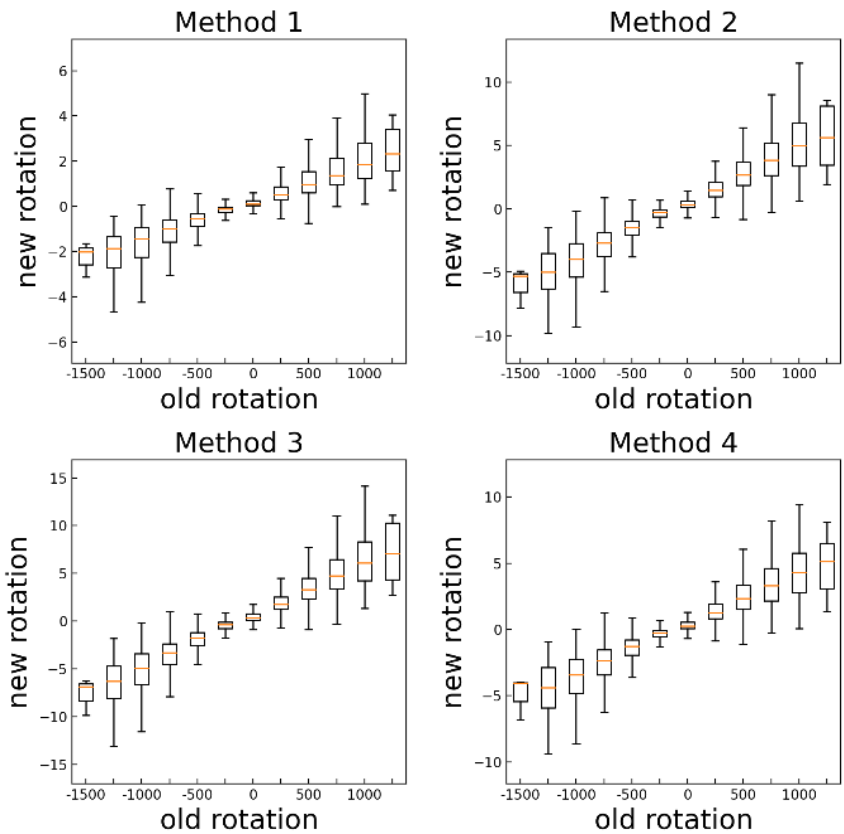

Figure 11. The relationship between old and new rotation concerning different methods
(Note: Box plot data are human current averaged over different day from 2022/3 to 2022/11, time period is from 5 am to 24 pm per 30 minutes, locations is about 30,000 nodes in the greater Tokyo area, sample size is the number of locations times time period, which is about 1,114,000)

Fig.12 shows the CDF of rotation calculated by different methods. It is meaningless to compare the rotation value calculated by different methods directly. Thus, we make the mean value of rotation calculated by all methods consistent and then compare. As a result, there is still not much difference between rotation calculated by different new methods with respect to distribution.

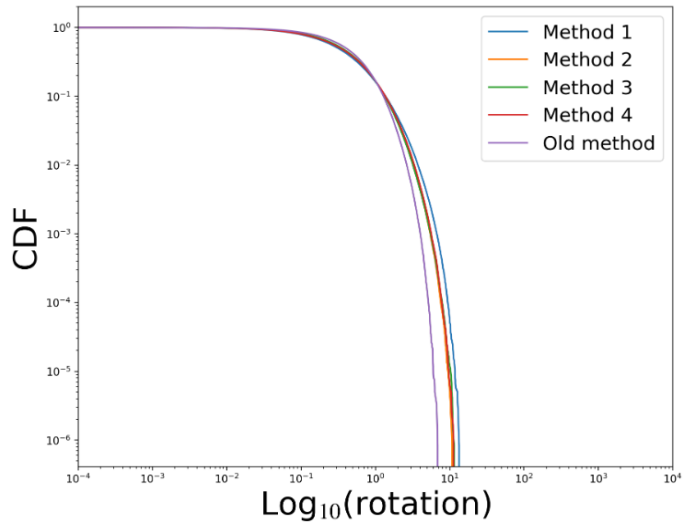

Figure 12. Comparison of CDF of rotation calculated in different methods

Moreover, we tried to plot the value of the loss function (Eq.(2)) of Adam concerning the different time periods in a day. As shown in Fig.13, the purple colour line (0 iterations) drops to the brown colour line after the 2000 iteration and drops to the pink one after the 10,000 iterations. The value of the loss function decrease, and the time series becomes steadier. We tried to compare the trend of the time series, and we found that the trend of our case (red colour line with rectangle) is close to the one after the 2000 iteration.

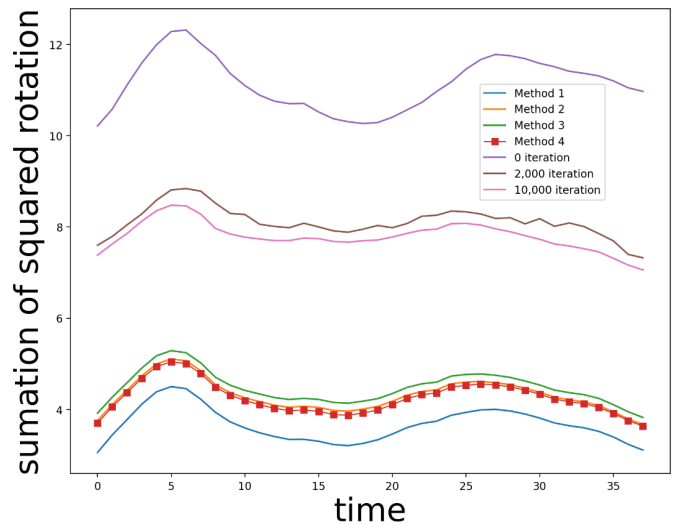

Figure 13. Comparison of loss function when calculating resistance concerning different methods (Note: line with rectangle is the method recommended in our study)

In Fig.14, rotation spatial distribution shows we also achieve removing rotation of electric field in electric circuit as much as possible in rush hour (left), and there is almost no rotation during other time period (right).

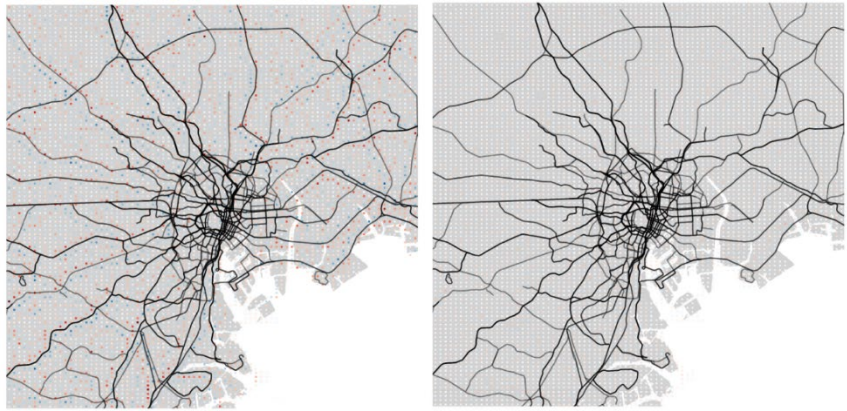

Figure 14. Rotation spatial distribution calculated by method recommended by us (Note: left is from 7:30 to 8:00 rush hour, right is from 13:30 to 14:00)

In Fig.15, when time period is fixed to 14:00 pm at which human current is minimum with a day, variant of current among different day keeps a linear relation with human conductivity, the inverse of human resistance, which is also consistant with the result claimed in the previous research, even though we changed the method to calculate resistance value.

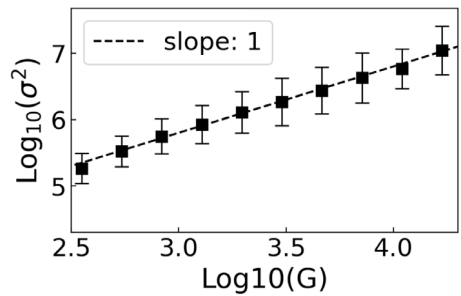

Figure 15. Relation between conductivity and variant of current

Overall, we consider that our new methods can replace the old method in the previous research to calculate human potential because it shows that new and old potential keep a well-linear relationship in Fig.4. And our new methods solved the limitation of the previous method, including high computational cost, results being not unique, hard to deploy and use due to fine-tuning of hyperparameters, and calculation error (outlier) in rural areas, as told in 2.2. Besides, there is not much difference between the property of new and old methods proposed new methods. Therefore, people can choose the suitable method to calculate resistance and human potential.

## 3. Route Generation Model

In this section, we will propose our recommended method (using current mean value) to determine resistance in 3.1 and show that there are three advantages to doing so. Then, we show that the result of RGM does not rely on the system's boundary in 3.2, so that people can select any observation area that is interested in generating the route. Finally, we will give more examples of the route generated by RGM in 3.3.

### 3.1 The method we recommend using to determine resistance

Section 2 showed there are not much difference between different methods to calculate resistance, including different quantile currents and mean current values. In this section, we recommend that using the mean absolute current value is the best way to calculate resistance. There are three reasons for it.

The first reason is that according to LLN (Law of Large Numbers), the mean value will converge to the expectation of the exponential distribution and become steadier and more robust with the increase of data. The resistance is supposed to reflect the infrastructure level so the value should be determined as robust as possible from the data.

The second reason is that we can implement real-time online updates for resistance value. Given the old mean current value $MI_{old}$ and the old number of samples $N_{old}$, when data of human current of the next day $I_{new}$ comes, the new mean current value $MI_{new}$ of a place can be updated instantly as follows.

$$MI_{new} = \frac{MI_{old}*N_{old}+I_{new}}{N_{new}} \quad (13)$$

where, $N_{new} = N_{old} + 1$

This method's time and space complexity both is $O(1)$ because we only need to maintain and update two variables $MI$ and $N$. While updating per cent quantile current requires maintaining an array when new data comes, sorting the array and finding a specific per cent quantile current value. The time complexity of sort algorithms is $O(n^2)$ or $O(n*log\ n)$ depend on different array properties. If the user maintains a sorted array, it still spends $O(n)$ to search a certain per cent quantile number or spends $O(log\ n)$ to search it by using binary search for an ordered array. Moreover, the space complexity of the latter is also $O(n)$. Therefore, using mean value can make the time and space complexity of real-time online update $O(1)$, which is independent of the size of system $n$ and more suitable to be applied to big data.

The third reason is that mean current has better interpretability for human resistance. In the previous research, optimization algorithm was like a black box, and it is hard to therotically parse the meaning or property of calculated human resistance. However, in our study, in the main text of this paper, we have the following Eq. (14-16):

$$I_{(d,t,L,\sigma)} = \frac{(v_{(d,t,L,\sigma)}*p_{(d,t,L)}+v_{(d,t,\sigma(L),\sigma)}*p_{(d,t,\sigma(L))})}{2} \quad (14)$$

$$R_{(L,\sigma)} := \frac{1}{\operatorname*{mean}_{d,t}\{|I_{(d,t,L,\sigma)}|\}} \tag{15}$$

$$\rho_{(L,\sigma)} = 1/\min_{dir}\{R_{(L=des,\sigma)}\} \tag{16}$$

Eq.(17) can be obtained by combining Eq.(14-16)

$$\rho_{(L,\sigma)} = \frac{1}{R_{(L,\sigma)}} = \operatorname*{mean}_{d,t}\{|I_{(d,t,L,\sigma)}|\} = \operatorname*{mean}_{d,t}\left\{\frac{(v_{(d,t,L,\sigma)}*p_{(d,t,L)} + v_{(d,t,\sigma(L),\sigma)}*p_{(d,t,\sigma(L))})}{2}\right\} \tag{17}$$

Where $\sigma \in \{\sigma_{+x}, \sigma_{+y}, \sigma_{-x}, \sigma_{-y}\}$

We explain why conductivity reflects infrastructure level in the following text. When the population $p_{(d,t,L)}$ in node $L$ is fixed, the faster people move in an area, the larger $v_{(d,t,L)}$ (mean average velocity for every people in node $L$, date $d$, time period $t$), means this area has better transportation tools. For example, according to the moving speed, transportation can be roughly classified into four classes: walking (below 10 km/h), bus (10 km/h – 80 km/h), subway and train (80 km/h – 100 km/h), and high-speed train (200 km/h – 250 km/h). Conversely, if the average moving speed of every people in a node $v_{(d,t,L)}$ is fixed, higher $p_{(d,t,L)}$ means higher population density that the node can transport, which is also reflecting the infrastructure level. (We determined the resistance value using the mean current value for the three reasons above)

### 3.2 Boundary selection of RGM

In the main text of our paper, we suggest that people can determine the number of nodes and the shape of the boundary in RGM arbitrarily, satisfying the following two conditions, 1) only one connected component in the electric circuit network, and 2) the network include the origin node and destination node. The first condition must be satisfied because a possible route must exist from the origin to the destination.

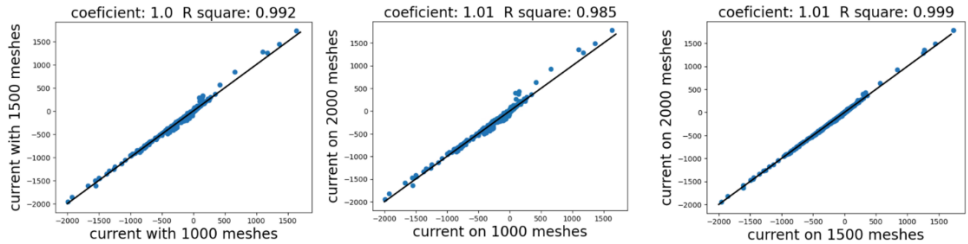

Figure 15. Relationship between different size of system. (Note: x-axis and y-axis are the comparisons between generated current value when network contains 1500, 2000, and 2500 nodes, respectively)

As shown in Fig.15, we make the observation area larger (change the network size) and run RGM (the origin and the destination are the same as the places mentioned in the body text). It is found that the solution of RGM in different network size keep a well-linear relationship (slope of linear regression is 1), which means the solution does not rely on the boundary that people set, and the boundary can be to any convenient shape for computating.

### 3.3 More examples of route generation

We run our RGM and found our model works well on other places on the map, as shown in Fig. (16-17). Our model generates recommended routes provided by Google Maps. In rural areas, generated roads will be relatively simple because people can only go the highway or the railway, which is close to 1-dimension. In urban areas, generated roads will be relatively complicated because there are too many available routes to the destination, which is close to 2-dimension.

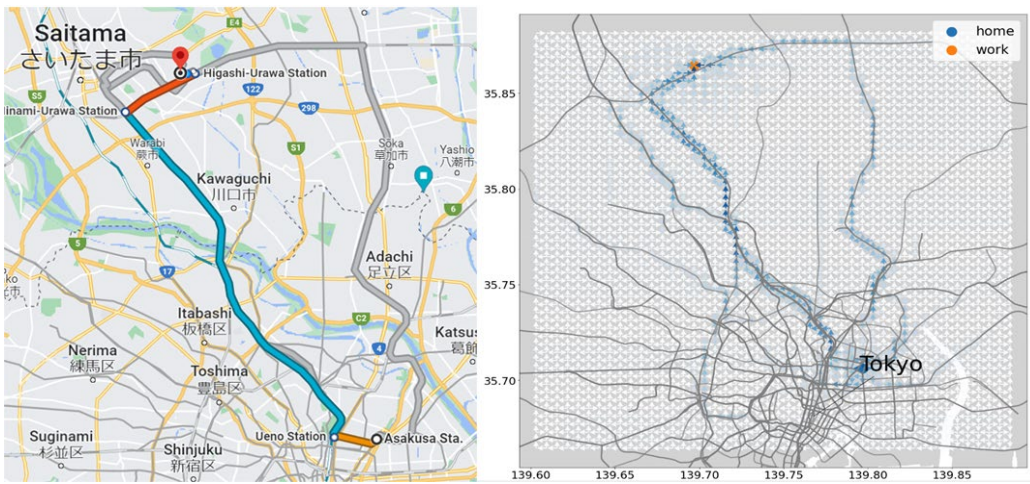

Figure 16. RGM example 1 (origin: Asakusa station, destination: Higashi-Urawa station, left picture cited from Google Map[2])

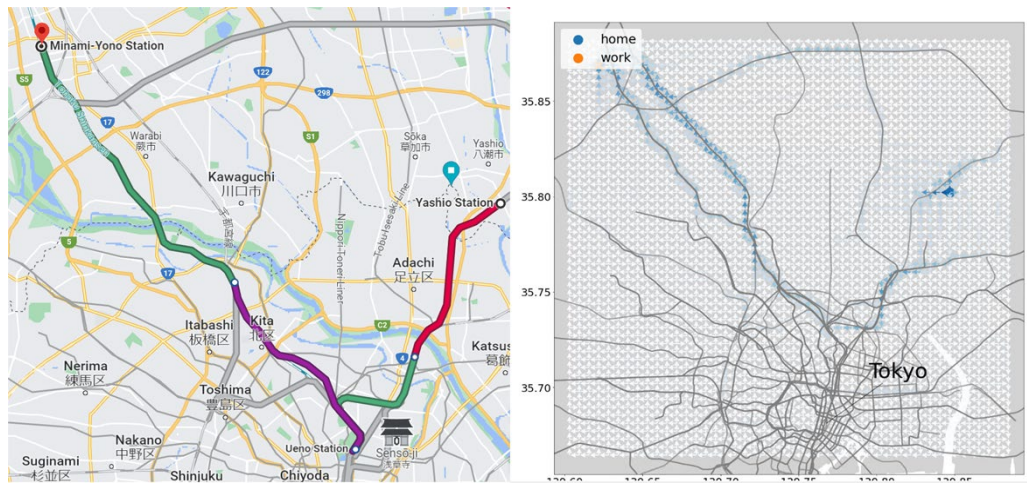

Figure 17. RGM example 2 (origin: Yashio station, destination: Minami-Yono station, left picture cited from Google Map[3])

As shown in Figure 18, when Tama monorail was blocked in the left of Tokyo, people will bypass that area to go to the destination.

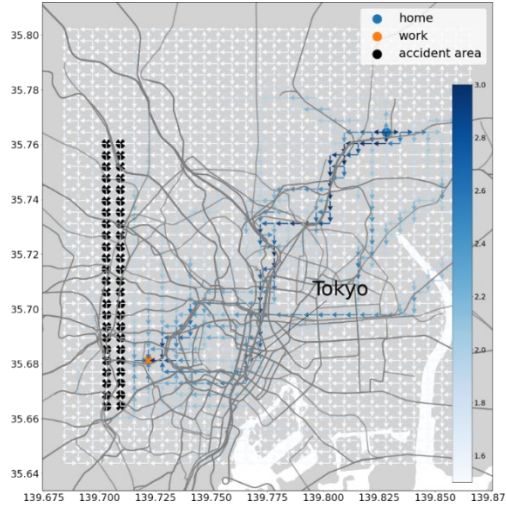

Figure 18. Route generation by RGM (grey line: railway line, black dots: accident area near Tama monorail, example 2, home: where people live, work: where people go to work)

## 4. Review of the method of calculating human potential

In the previous research[1], a detailed way to calculate electric potential on a map is not shown clearly. Therefore, as a supplement, we want to propose our procedure to calculate the value of potential in detail.

Firstly, we need to calculate the charge $Q_{(d,t,L)}$, the divergence of voltage, at date $d$, time period $t$ and location $L$.

$$Q_{(d,t,L)} = div(I_{d,t,L} * R_L) = \sum_{i=1}^{4} I_{(d,t,L,\sigma_i)} * R_{(L,\sigma)} \qquad (18)$$

The map's mesh number is fixed when the observed data are given. In this case, we can view each mesh as a node in the 2-dimensional square lattice and define the adjacent matrix for each node. First, find the network's largest connected component by Breadth-First Search algorithm. Then, we define the adjacent matrix:

$$A_{i,j} = \begin{cases} 1, & node\ i\ is\ adjacent\ to\ j \\ 0, & else \end{cases} \qquad (19)$$

Moreover, we define the diagonal matrix, reflecting how many nodes are adjacent nearby, for the node on the sea boundary (sea boundary is defined by using land use data that the sea ratio of a node above 95% will be viewed as a sea area. The data was statistics by the Japan government):

$$D_{i,j} = \begin{cases} \sum_j A_{i,j}, & case\ 1 \\ 4, & case\ 2 \\ 0, & case\ 3 \end{cases} \qquad (20)$$

where, $case\ 1: i = j\ and\ node\ i\ is\ adjacent\ to\ sea\ area$
$case\ 2: i = j\ and\ node\ i\ is\ on\ the\ other\ area$

$$case\ 3: i \neq j$$

Theoretically, a boundary condition that $\emptyset = 0$ near the land boundary needs to be set, but practically, we only need to solve human potential $\emptyset_{d,t,l}$ at day $d$, time period $t$, location $L$ by solving the following formula. The number of variables and linear equations is the number of nodes (locations).

$$(D_{i,j} - A_{i,j}) * \begin{pmatrix} \emptyset_{d,t,1} \\ ... \\ \emptyset_{d,t,L} \end{pmatrix} = \begin{pmatrix} div(I_{d,t,1} * R_1) \\ ... \\ div(I_{d,t,L} * R_L) \end{pmatrix} \quad (21)$$

Matrix $(D_{i,j} - A_{i,j})$ also called Laplacian matrix $\Delta_{i,j}$ in a network. The time complexity of BFS is $O(V + E)$, where $V$ is the number of vertices and $E$ is the number of edges. In our case, $V$ equals $L$, and $E$ approximately equals $4 * L$ because it is a lattice network. Therefore time complexity of BFS is $O(4 * L + L) \rightarrow O(L)$.

The time complexity of calculating charge and solving linear equations is $O(L * d * t)$ and $O(L^3)$ respectively, therefore the total time complexity of calculating potential is $O(L) + O(L * d * t) + O(L^3) \rightarrow O(L^3)$. Similar to the discussion part in main text, we can reduce the computational cost to $O(L^{2.33})$ practically, when the coefficient matrix is very sparse[5].

## 5. Brief introduction of our open-source code

Users can download open source computer programming language, called Python 3, to use our code to conduct research about human flow pattern. The code related to RECM will be provided as a Python script file in the following URL: https://github.com/Zhihua-Zhong/Revised-Electric-Circuit-Model. The method to calculate human current, human resistance and other variables has been encapsulated into Python functions that can be called and customized by other programmers and users. People with input data that fulfils the standard input data format shown in Fig. 1 can use our codes to implement RECM directly. If not, a simple format transformation is required beforehand. Regarding the development environment, python 3.8.5 with third-party dependencies, including pandas, geopandas, numpy, matplotlib, and scipy are required. More details and hand-to-hand tutorials can be found on our GitHub through the above URL.

In order to increase computing efficiency, multi-process calculation functions were also developed to enable users with powerful computers to use multi CPUs to compute the result simultaneously. Specifically speaking, because the calculation of human current and resistance is mutually independent at different nodes and independent on different days, the calculation on different spaces and times can be separated and paralleled to the different CPUs.

Our open-source code will be updated over time and new function will be gradually online to respond to other researchers' new demands. If there is any query, please do not hesitate to contact us through the contact information revealed in the above website.